\documentclass[reprint,superscriptaddress,aps,prx]{revtex4-2}

\usepackage{graphicx}
\usepackage{circledsteps}
\usepackage{amssymb}
\usepackage{amsmath}
\usepackage{epsfig}
\usepackage{chemarr}
\usepackage{url}
\usepackage{natbib}
\usepackage{dcolumn}
\usepackage{bm}
\usepackage{xcolor}
\usepackage{braket}

\usepackage[
colorlinks=true,
linkcolor=blue,
citecolor=blue,
urlcolor=blue
]{hyperref}

\def\be{\begin{equation}}
\def\ee{\end{equation}}
\def\bmu{\begin{multline}}
\def\bea{\begin{eqnarray}}
\def\eea{\end{eqnarray}}
\def\nn{{\hat{\mathbf{n}}}}
\def\zz{{\hat{\mathbf{z}}}}
\def\rr{{\mathbf{r}}}
\def\xx{{\mathbf{x}}}
\def\XX{{\mathbf{X}}}

\def\Dr{D_\mathrm{r}}

\begin{document}

\title{Robust Topologically Protected Edge Transport in Doubly Chiral Active Particles}

\author{Tristan Edwards}
\thanks{Equal contribution.}
\address{Department of Physics and Astronomy, University College London, London WC1E 6BT, United Kingdom}
\author{Maxim Nikolaev}
\thanks{Equal contribution.}
\address{Department of Physics and Astronomy, University College London, London WC1E 6BT, United Kingdom}
\author{Jaime Agudo-Canalejo}
\email{j.agudo-canalejo@ucl.ac.uk}
\address{Department of Physics and Astronomy, University College London, London WC1E 6BT, United Kingdom}

\begin{abstract}
Using theory, simulation, and experiment, we introduce a new class of active particle which we term doubly chiral active Brownian particles (dcABPs), which show robust topologically protected transport along boundaries without backscattering at corners. Their double chirality stems from the coexistence of an intrinsic angular velocity, which can cause rotation independently of translation, and a translation-rotation coupling inducing cross-alignment to the instantaneous velocity, which causes rotation only concomitantly with translation. A mechanically detailed model shows that the latter effect can arise from an asymmetric friction distribution in the direction perpendicular to the self-propulsion direction. We show that topologically protected modes emerge when the two sources of chirality have opposite sign and  the intrinsic rotation is weaker than the translation-rotation coupling. In the deterministic limit, we characterize the emergence of these modes not only along straight boundaries, but also along curved boundaries and during interparticle interactions. We provide a proof-of-principle experimental realization by building a doubly chiral vibrobot. While setting the work into context, we moreover show that the topologically protected boundary-induced transport of dcABPs stands in contrast to the edge currents observed for simple chiral ABPs, which we demonstrate are not associated with boundary-induced transport, as well as to those observed for chiral active rods or self-aligning chiral ABPs, which we show to be associated with boundary-induced transport but to backscatter at corners, implying lack of topological protection.
\end{abstract}

\maketitle

\section{Introduction}

Many works in recent years have linked the dynamics of chiral active matter to the potential existence of edge currents along boundaries or interfaces \cite{Shankar2022,vanZuiden2016ActiveSpinners,Soni2019OddFreeSurface,Yang2020RobustBoundaryFlow,Jamali2018ActiveCircularBoundaries,metzger2026equation,wang2026edgecurrentsshapecondensates,PhysRevX.12.041017,PhysRevX.14.041006}, in analogy with how cyclotron orbits undergone by electrons in the presence of a magnetic field cause skipping orbits and ultimately Hall currents in a two-dimensional material \cite{Halperin1982EdgeStates,Buttiker1988AbsenceBackscattering}.
Edge currents have been reported to exist in chiral active liquids made of self-spinning particles \cite{vanZuiden2016ActiveSpinners,Soni2019OddFreeSurface,Yang2020RobustBoundaryFlow,
Petroff2023DensityMediatedSpin} or living cells \cite{PhysRevX.12.041017,PhysRevX.14.041006}, where they arise as a collective effect. They have also been reported in chiral Active Brownian Particles (cABPs, also known as circle swimmers), where steady state probability currents are observed near boundaries even at the single particle level \cite{Caprini2019ActiveChiralConfinement,caprini2025activethermodynamicsinertialchiral,metzger2026equation}. Edge currents have also been clearly demonstrated for single chiral active rods (or similarly, self-aligning cABPs), where early theory and experiments on chiral active motion \cite{vanTeeffelen2008,teeffelen2009clockwise,Kummel2013CircularMotion} as well as more recent works \cite{Deblais2018BoundariesControl,kant2025edge,carrillomora2025depinningactivatedmotionchiral} have shown the existence of sliding modes along straight or smoothly curved boundaries.

Characterizing edge currents and associated edge-induced transport is no easy task. One key issue is that the existence of edge currents may not necessarily be associated with the existence of edge-induced transport, in analogy with the distinction between magnetization currents and transport currents in electronic transport
\cite{Cooper1997Thermoelectric,PhysRevLett.107.236601}. A second key issue is that edge currents may or may not be topologically protected, i.e.~robust against deformations of the boundaries of the system. In particular, topological protection ensures that edge currents do not \emph{backscatter}, i.e.~reflect backwards into the bulk, in the presence of boundary imperfections or defects \cite{Buttiker1988AbsenceBackscattering,HasanKane2010TopologicalInsulators,
Lu2014TopologicalPhotonics}.

An example of chiral active matter that shows topologically protected boundary-induced transport even at the single particle level still has not been unambiguously demonstrated. Intriguingly, the same is not true for discrete-space, two-dimensional lattice models describing out-of-equilibrium classical stochastic dynamics, where topologically protected edge currents have been clearly demonstrated \cite{Tang2021TopologyProtectsChiralCurrents,AgudoCanalejo2025}. Considering that such lattice models can often represent a discretized version of continuum single-particle dynamics, it is worth considering whether topologically protected boundary-induced transport can be achieved for single active particles.

Here, we show that such an active particle can indeed be constructed, not only in theory but also in practice. To set the stage, we first consider in detail whether existing models of chiral active particles display topologically protected boundary-induced transport at the single particle level (Section~\ref{sec:stateoftheart}). Throughout this work, we define topologically protected  transport operationally as boundary-localized directed motion that persists under continuous boundary deformations and does not backscatter into the bulk at corners or local defects. We show that single cABPs (circle swimmers) do not undergo boundary-induced transport, even if they display non-zero probability currents near boundaries at steady state. In turn, we confirm that single chiral active rods (or, similarly, self-aligning cABPs) do undergo boundary-induced transport, but we show that they backscatter at inside corners, implying lack of topological protection.

Taking inspiration from the lattice model mentioned above \cite{Tang2021TopologyProtectsChiralCurrents}, we then introduce and characterize Doubly Chiral Active Brownian Particles (dcABPs) as a model that demonstrates topologically protected boundary-induced transport, without any backscattering at corners (Section~\ref{sec:dcabps}). The two sources of chirality correspond to an intrinsic rotation, and a translation-rotation coupling that causes cross-alignment to the instantaneous velocity. In the deterministic limit, we characterize the boundary-sliding modes for both straight and curved boundaries, as well as the spinning modes when two particles interact with each other.

Finally, we consider the challenge of building a dcABP in practice (Section~\ref{sec:practical}). Through a mechanically detailed theoretical model, we show that the required translation-rotation coupling arises naturally from an asymmetric friction distribution in the direction perpendicular to the direction of motion, and how the mechanically detailed model reduces to the dcABP model in a suitable limit. We finish by demonstrating how a doubly chiral vibrobot showing sliding modes with no backscattering can be built using simple parts.

\section{Evaluating the state of the art \label{sec:stateoftheart}}

\subsection{Absence of edge transport in simple chiral ABPs (cABPs)}

\subsubsection{Single overdamped cABPs}

The standard, widely studied model for chiral ABPs (cABPs, also known as circle swimmers) corresponds to the dynamics
\begin{eqnarray}
    \dot{\rr} & = & v \nn + \mu \mathbf{F} \label{eq:rdyn} \\
    \dot{\theta} & = & \omega + \sqrt{2\Dr} \eta  \label{eq:thetadyn_cABP}
\end{eqnarray}
where $\rr = (x,y)$ is the position, $\nn = (\cos\theta,\sin\theta)^t$ is the orientation, $\bf{F}$ is the external force (due to boundaries or other particles), $v$ is the self-propulsion speed, $\mu$ is a mobility (inverse friction), $\omega$ is the intrinsic angular velocity (source of the chirality), $\Dr$ is the rotational diffusion coefficient, and $\eta(t)$ a white noise satisfying  $\langle \eta(t) \rangle = 0$ and $\langle \eta(t) \eta(t') \rangle = \delta(t-t')$. In the bulk, away from boundaries, and for sufficiently low noise, such particles will typically undergo chiral orbits with radius $R_\mathrm{bo}=v/|\omega|$.

\begin{figure*}
	\centering
	\includegraphics[width=1\textwidth]{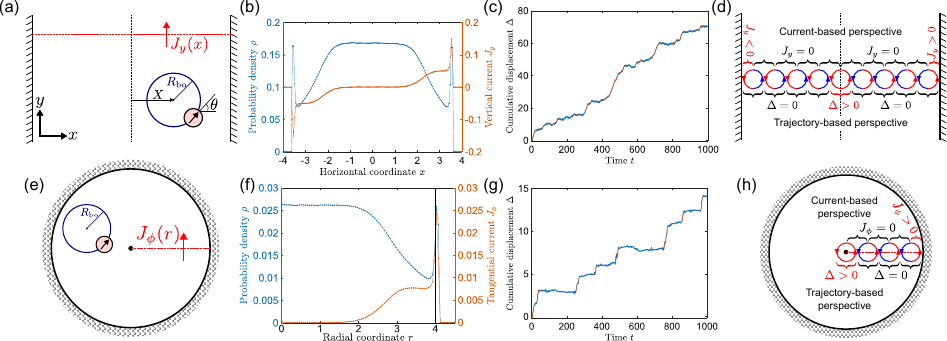}
	\caption{Absence of edge transport in cABPs. Schematics of the (a) channel and (e) circular arena geometries. (b,f) Steady state probability density (blue) and vertical current (red) along the horizontal (channel) or radial (arena) coordinates, showing local net currents at the boundaries with the same counterclockwise chirality as the bulk orbits. (c,g) Evolution of the signed cumulative displacement in the vertical (channel) or tangential (arena) direction over time in the same simulations, which shows net drift (in the positive direction) only when the particle is within one bulk orbit radius from the channel or arena center (red segments). (d,h) Duality in the apparent origin of the integrated current $\Phi$. When integrating steady state currents in space, it wrongly appears as if the current arises due to boundary-induced transport. A trajectory-based perspective reveals that boundaries do not induce transport, and $\Phi$ arises from bulk orbits near the center of the channel or arena. For parameter values used in simulations, see Appendix~\ref{app:simulationdetails}.
	}\label{fig:fig1}
\end{figure*}

Edge currents at boundaries have been frequently reported for such cABPs, even at the single particle level. Indeed, experiments, simulations, and theory measuring or calculating such steady state averaged currents appear to show a nonzero current near hard boundaries \cite{metzger2026equation}, e.g.~a nonzero $J_y(x)$ in a channel geometry where the confinement is along $x$ [Fig.~\ref{fig:fig1}(a,b)]. These edge currents along inner boundaries are seen to have the same chirality as the bulk orbits of the particles, i.e.~resulting in upwards currents on the right boundary and downwards currents on the left boundary for counterclockwise bulk orbits ($\omega>0$), and vice versa for clockwise bulk orbits. However, we argue here that such currents do not correspond to real boundary-induced transport, in analogy with the distinction between magnetization currents and transport currents in electronic transport
\cite{Cooper1997Thermoelectric,PhysRevLett.107.236601}. 

Let us consider first the following simple argument for a channel geometry. In such a geometry, the boundary exerts forces only along $x$, and thus the dynamics of the $y$ coordinate is simply $\dot{y}=v \sin \theta$ [Eq.~\ref{eq:rdyn}]. Because the dynamics of $\theta$ are fully autonomous [Eq.~\ref{eq:thetadyn_cABP}], then the $y$ dynamics of the particle are necessarily independent of whether the channel boundaries are present or absent. This in itself is a definitive argument against the existence of boundary-induced particle transport in a channel geometry. A simulation example of a cABP interacting with a straight boundary is shown in Movie S1.

What is, then, the significance of the nonzero vertical current $J_y(x)$ near the channel boundaries? Even more puzzlingly, what about the nonzero integrated vertical current $\Phi = \int_{0}^{L/2} J_y(x)\mathrm{d}x = - \int_{-L/2}^0 J_y(x)\mathrm{d}x$ (in a channel of width $L$)? The integrated current $\Phi$ being nonzero certainly seems to imply the existence of boundary-induced transport, and $\Phi$ was calculated in Ref.~\citenum{metzger2026equation} to be generically given by
\begin{equation}
    \Phi=\frac{ \rho_\mathrm{b} v^2 \omega}{2(\omega^2+\Dr^2)} \underset{\Dr \ll \omega}{\simeq} \frac{ \rho_\mathrm{b} v^2 }{2\omega} . \label{eq:intcurrent}
\end{equation}
where $\rho_\mathrm{b}$ is the bulk probability density, and the second result corresponds to the limit of small rotational noise.

First, let us note that the integrated current $\Phi$ can equivalently be expressed as the (appropriately signed) time-averaged velocity of the particle along the vertical direction, namely $\Phi = \lim_{t \to \infty} \Delta(t)/t$ where $\Delta (t)$ is the cumulative signed vertical displacement up to time $t$, which for the channel geometry is $\Delta(t)=\frac{1}{2} \int_0^t \mathrm{sgn}(x)\dot{y} \mathrm{d}t'$. Now, consider the behavior of $\Delta(t)$ over time for a single particle [Fig.~\ref{fig:fig1}(c)]. When it is far from the channel center, $|x|>R_\mathrm{bo}$, the particle describes bulk orbits and the cumulative vertical displacement $\Delta(t)$ only oscillates, while undergoing some unbiased diffusion due to nonzero rotational diffusion $\Dr$, but it has no net drift. This is true even when particles collide against the channel boundaries and their trajectories are thus distorted. Only when particles are near the channel center, with $|x|<R_\mathrm{bo}$ then both halves of a bulk orbit can contribute constructively to the cumulative displacement $\Delta(t)$ after a full orbit, generating a net increase of $\Delta(t)$. Thus, from a particle-centric point of view, the nonzero integrated current $\Phi$ is due only to bulk orbits near the channel center, far from the channel boundaries. 

This argument can be made quantitative, so as to recover the exact expression for the integrated current $\Phi$ [Eq.~\ref{eq:intcurrent}] only from the bulk dynamics, without any reference to the existence of boundaries. We describe here the simpler derivation in the limit of small rotational noise ($\Dr \ll \omega$), which is purely geometrical, and refer to Appendix~\ref{app:withnoise} for the general case. In the limit of low noise, particles describe almost perfect bulk orbits, with the small noise only causing diffusion of the center of these orbits so that the system remains ergodic. Let $X$ be the signed distance from the channel centerline to center of a bulk orbit. The full orbit is parametrized by the particle orientation $\theta \in [0,2\pi)$. In particular, the particle position satisfies $x=X+R_\mathrm{bo} \sin \theta$, and the vertical velocity $\dot{y}=v \sin \theta$. The integrated current can then be calculated as
\begin{eqnarray}
    \Phi = \frac{1}{2} \rho_\mathrm{b} \frac{1}{2\pi} \int \mathrm{d}X  \int_0^{2\pi} \mathrm{d}\theta \, \mathrm{sgn}(X+R_\mathrm{bo}\sin\theta) v \sin \theta.
\end{eqnarray}
Note that, when $|X|>R_\mathrm{bo}$, then the sign function inside the integral is constant, and the innermost integral evaluates to zero as $\int_0^{2\pi}\sin\theta=0$. Only when $|X|<R_\mathrm{bo}$ and the orbit intersects with the channel center we get a nonzero contribution from the inner integral, resulting in
\begin{eqnarray}
    \Phi = \frac{\rho_\mathrm{b}v}{2\pi} \int_{-R_\mathrm{bo}}^{R_\mathrm{bo}} \mathrm{d}X \left(  \int_{\pi/2}^{\theta_*} \mathrm{d}\theta \, \sin \theta - \int_{\theta_*}^{3\pi/2} \mathrm{d}\theta \, \sin \theta \right)
\end{eqnarray}
with $\sin \theta_* \equiv - X/R_\mathrm{bo}$. This evaluates to
\begin{eqnarray}
    \Phi = \frac{\rho_\mathrm{b} v R_\mathrm{bo}}{2} = \frac{\rho_\mathrm{b} v^2}{2\omega}
\end{eqnarray}
and coincides with Eq.~\ref{eq:intcurrent}. Importantly, we obtained this result from the dynamics at the center of the channel without any reference to the existence of boundaries, in contrast with Eq.~\ref{eq:intcurrent}, which was obtained in Ref.~\citenum{metzger2026equation} by integration of $J_y(x)$ near a boundary.

A similar argument can be made for a cABP confined inside a circular arena, where it becomes beautifully topological. Now, the current of interest is that along the tangential direction $J_\phi(r)$, with $r$ and $\phi$ the radial and angular coordinates. The integrated tangential current is $\Phi = \int_0^R J_\phi(r) \mathrm{d}r$, where $R$ is the radius of the arena, or equivalently $\Phi = \lim_{t \to \infty} \Delta(t)/t$ with $\Delta(t) = \frac{1}{2\pi}\int_0^t \dot{\phi}\mathrm{d}t'$. This integrated current measures the net number of counterclockwise crossings through a radial cut along the arena per unit time [Fig.~\ref{fig:fig1}(e)]. As for the channel, we find that while the local current $J_\phi(r)$ is nonzero only at the boundaries [Fig.~\ref{fig:fig1}(f)], the cumulative displacement $\Delta(t)$ grows only when the particle is near the center of the arena [Fig.~\ref{fig:fig1}(g)]. Let us consider again the limit of small rotational noise, where particles undergo bulk orbits of radius $R_\mathrm{bo}$ (see Appendix~\ref{app:withnoise} for the general case). First, note that orbits that do not enclose the center of the arena do not contribute any net crossings. Orbits that do enclose the center, on the other hand, contribute one crossing per period of the orbit, which is $T=2\pi/\omega$. The centers of such orbits must lie within a circle of radius $R_\mathrm{bo}$ around the center of the arena. Thus, the integrated current is given by 
\begin{eqnarray}
    \Phi = \frac{\rho_\mathrm{b} \pi R_\mathrm{bo}^2}{T} = \frac{\rho_\mathrm{b} v^2}{2\omega} \label{eq:intcurrent_arena}
\end{eqnarray}
where we have again recovered the same result, purely from bulk orbits without reference to the boundaries of the arena.

The above arguments demonstrate that single overdamped cABPs do not experience boundary-induced transport, despite a nonzero local current ($J_y(x)$ or $J_\phi(r)$) near the boundaries and a nonzero integrated current ($\Phi$) along an appropriate cut perpendicular to the boundary. Two completely different (and even seemingly contradictory) but ultimately equivalent ways of calculating the integrated current, one based on integrating local currents near the boundary \cite{metzger2026equation}, the other based on integrating particle trajectories far from the boundary, highlight an intriguing bulk-boundary duality [Fig.~\ref{fig:fig1}(d,h)]. From the perspective of local currents, particle orbits with different centers cancel each other out, everywhere except near boundaries, as there aren't any orbits with center ``inside'' the boundary to cancel currents tangential to the boundary. On the other hand, from a trajectory-based perspective, chiral orbits show no net displacement after a full orbit and thus do not contribute to the integrated current anywhere except near the center of the channel or arena. The duality highlights that, while in a technical sense there are indeed nonzero (integrated) boundary currents in this system, these currents do not imply boundary-induced transport. This result appears analogous to the distinction between magnetization currents and transport currents in electronic transport \cite{Cooper1997Thermoelectric,PhysRevLett.107.236601}.

\subsubsection{The role of inertia}

Our main focus here is on overdamped dynamics. However, one may wonder whether true boundary-induced transport may be recovered for underdamped cABPs, where Eq.~\ref{eq:rdyn} is replaced by $\tau \ddot{\rr}  =   - \dot{\rr} + v \nn + \mu \mathbf{F}$ with $\tau \equiv \mu m$ the inertial relaxation time, while Eq.~\ref{eq:thetadyn_cABP} remains unchanged. Importantly, the same simple argument for the channel geometry presented above still applies in this case: the dynamics of the vertical coordinate are simply $\tau \ddot{y}  =   - \dot{y} + v \sin\theta$, and the dynamics of $\theta$ are autonomous. Therefore, the vertical dynamics of the particle are unaffected by the presence of the straight boundaries, and there cannot be any boundary-induced transport. Boundary-induced transport may still exist for underdamped cABPs in circular confinement, as in this case the radial and tangential dynamics become coupled by the curved boundary, potentially leading to momentum transfer.

\subsubsection{The role of interparticle interactions.}

Going back to overdamped cABPs, it is worth considering whether boundary-induced transport does re-emerge as a collective phenomenon for many-particle systems, thanks to interparticle interactions. Consider again a channel geometry, now with $N$ particles. The dynamics of the vertical coordinate of particle $i$ can be written as $\dot{y}_i = v \sin \theta_i + \mu \sum_{j\neq i} F_{ij}^y$, where $F_{ij}^y$ is the force exerted by $j$ on $i$. The dynamics of the orientation $\theta_i$ remain autonomous, with $\dot{\theta}_i = \omega + \sqrt{2\Dr} \eta_i$. Generalizing the trajectory-based expression for the integrated current $\Phi$ introduced above to the case with many particles, we may write
\begin{equation}
    \Phi = \frac{1}{N} \lim_{t \to \infty} \frac{1}{t} \frac{1}{2} \int_0^t \left( \sum_{i=1}^N \mathrm{sgn}(x_i)\dot{y}_i \right) \mathrm{d}t'. \label{eq:collective1}
\end{equation}

Let us now evaluate the integrand in Eq.~\ref{eq:collective1}. Substituting the value of $\dot{y}_i$, manipulating the summation indices, and using Newton's 3rd law ($F_{ij}^y = -F_{ji}^y$), we obtain
\begin{eqnarray}
     \sum_{i=1}^N \mathrm{sgn}(x_i)\dot{y}_i & = & v \sum_{i=1}^N \mathrm{sgn}(x_i)\sin\theta_i + \\
     & & \mu \sum_{i=1}^N \sum_{j=i+1}^N \left[ \mathrm{sgn}(x_i) - \mathrm{sgn}(x_j) \right] F_{ij}^y.   \nonumber \label{eq:collective2}
\end{eqnarray}
The first term on the r.h.s.~is associated with the self-propulsion dynamics. Note that, because of the autonomous dynamics of $\theta_i$, the integral of $\sin \theta_i$ over a rotation period vanishes on average. As in the single particle case, the only trajectories that contribute to the integrated current are those that straddle the centerline of the channel, which do not correspond to true boundary-induced transport. The second term on the r.h.s.~is associated with the interparticle interactions. The term in square brackets implies that interactions between pairs of particles that lie on the same side of the channel do not contribute to the integrated current. Only interactions between pairs of particles that straddle the centerline of the channel (with each particle on a different side) contribute to the integrated current. This suggests that, at least at low densities far from a percolation transition where force chains could propagate across the whole channel, contributions of interparticle forces to the integrated current do not reflect true boundary-induced transport.

\subsection{Backscattering of chiral active rods}

We now turn our attention to chiral active rods or, analogously, self-aligning chiral active particles \cite{baconnier2025self}. For these two types of particles, boundaries can exert a torque (or a torque-like influence) on the angular dynamics, which in turn leads to true boundary-induced transport \cite{kant2025edge}. In particular, let us now consider the dynamics
\begin{eqnarray}
    \dot{\rr} & = & v \nn + \mu \mathbf{F} \label{eq:rdyn2} \\
    \dot{\theta} & = & \omega + \zeta (\nn \times \dot{\rr})\cdot \zz  +  \sqrt{2\Dr} \eta  \label{eq:thetadyn_rod} \\
    & = & \omega + \zeta \mu (\nn \times \mathbf{F})\cdot \zz  +  \sqrt{2\Dr} \eta \nonumber
\end{eqnarray}
where the only difference with respect to (\ref{eq:rdyn}--\ref{eq:thetadyn_cABP}) is the addition of the term with the parameter $\zeta>0$, which controls the anisotropy of the rod or equivalently the strength of self-alignment.

\begin{figure*}
	\centering
	\includegraphics[width=1\textwidth]{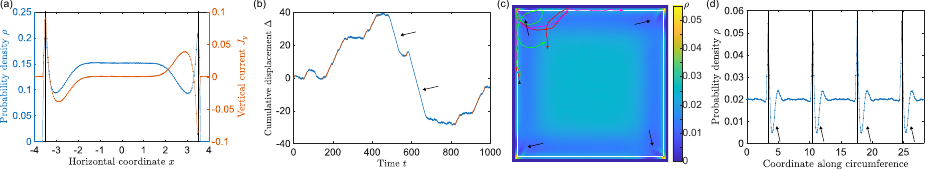}
	\caption{Backscattering of chiral active rods or self-aligning cABPs. (a) Steady state probability density (blue) and vertical current (red) along the horizontal coordinate in channel geometry, showing local currents with opposite chirality to bulk orbits at the boundaries. (b) Evolution of the signed cumulative displacement in the vertical direction over time in the same simulation, which shows net drift in the positive direction when the particle is within one bulk orbit radius from the channel centerline (red segments) and in the negative direction when the particle enters a transport-inducing sliding mode along the boundaries (black arrows). (c) Steady state probability density in square confinement. Three examples of particle trajectories, starting from the middle of the left boundary, are overlaid, and show clockwise boundary-sliding followed by backscattering as soon as the particle encounters the top left corner. Signatures of this backscattering are seen in the steady state probability density (black arrows). (d) Steady state probability density along the circumference of the system (just outside the boundary, in clockwise direction), with loss of probability (black arrows) just after each corner (vertical black lines) signalling backscattering. For parameter values used in simulations, see Appendix~\ref{app:simulationdetails}.
	}\label{fig:fig2}
\end{figure*}

It has been shown before that, in the deterministic limit $\Dr=0$, the dynamics given by (\ref{eq:rdyn}) and (\ref{eq:thetadyn_rod}) can have a stable sliding mode along straight and curved boundaries \cite{kant2025edge}. Indeed, let us consider without loss of generality a straight vertical boundary that confines a chiral active rod to the left hand side. Looking for a fixed point of the form $\dot{x}=\dot{\theta}=0$, we find that there is a critical angle $\theta_*$ satisfying $\sin(2\theta_*) = - \frac{2\omega}{\zeta v}$ where the particle orientation becomes locked. Such a critical angle exists as long as $\omega \neq 0$, $\zeta>0$, and $\left| \frac{2\omega}{\zeta v} \right|<1$. In this locked orientation, the particle slides along the boundary with a speed $v_* = v \sin\theta_*$.

In contrast to the simple cABP above, the edge currents resulting from this sliding mode have opposite chirality to the bulk orbits: for $\omega>0$ (counterclockwise orbits), we find $-\pi/2<\theta^*<0$, so that the particle is sliding downwards along a boundary on the right, and correspondingly upwards along a boundary on the left (in circular confinement, the sliding mode would be clockwise \cite{kant2025edge}). This opposite chirality of bulk orbits and edge currents is analogous to that seen for cyclotron orbits and skipping orbits (edge currents) in the quantum Hall effect \cite{Halperin1982EdgeStates}.

Stochastic simulations in a channel geometry confirm these expectations [Fig.~\ref{fig:fig2}(a)], showing edge currents with opposite chirality to the bulk orbits. Additionally, monitoring of the signed cumulative displacement $\Delta(t)$ shows that it systematically drifts in the negative direction when the particles are sliding against the boundaries (Fig.~\ref{fig:fig2}(b), indicated by black arrows), implying true boundary-induced transport in this case. This negative-sign transport-associated contribution to the integrated edge current $\Phi$ competes with the positive-sign `spurious' contribution due to bulk orbits near the channel center (Fig.~\ref{fig:fig2}(b), red segments). 

However, a topologically protected edge current should be robust to deformations of the shape of the boundary. In particular, topologically protected edge currents can avoid obstructions and propagate past defects such as sharp corners. If such obstacles obstruct the edge current and send particles back into the bulk, i.e.~induce \emph{backscattering}, then the current is not topologically protected \cite{Buttiker1988AbsenceBackscattering,HasanKane2010TopologicalInsulators,
Lu2014TopologicalPhotonics}.

Simulation of the dynamics given by (\ref{eq:rdyn}) and (\ref{eq:thetadyn_rod}) in square confinement shows that interior corners backscatter chiral active rods into the bulk (see Fig.~\ref{fig:fig2}(c) and Movie S2). Particles first become arrested at the corner, followed by continuous rotation in the direction favoured by the angular velocity $\omega$, until they are kicked out of the corner and begin to describe bulk orbits in its vicinity. This behaviour can also be observed in the experimental videos of Ref.~\citenum{kant2025edge}, and can be understood from the dynamics described by (\ref{eq:thetadyn_rod}). Indeed, when the particle comes into contact with and pushes into two perpendicular boundaries at a corner, it moves neither horizontally nor vertically. Thus, the self-alignment (or torque) term in (\ref{eq:thetadyn_rod}) vanishes and the particle simply rotates with angular velocity $\omega$. This continues until the particle has turned to point back in the direction that it came from, and is once again pushing only against one boundary. However, at this point the particle's orientation is no longer within the basin of attraction of the critical (locked) orientation $\theta_*$, and thus scatters into the bulk.

Backscattering results in a significant reduction of the edge current immediately after the corner, which picks back up until it gets backscattered again at the next corner [Fig.~\ref{fig:fig2}(d)]. Thus, although chiral active rods (or self-aligning cABPs) do undergo boundary-induced transport along straight or smoothly curved boundaries (e.g.~in circular confinement \cite{kant2025edge}), these edge states are not topologically protected against backscattering, and break down in the presence of sharp interior corners.

\begin{figure*}
	\centering
	\includegraphics[width=1\textwidth]{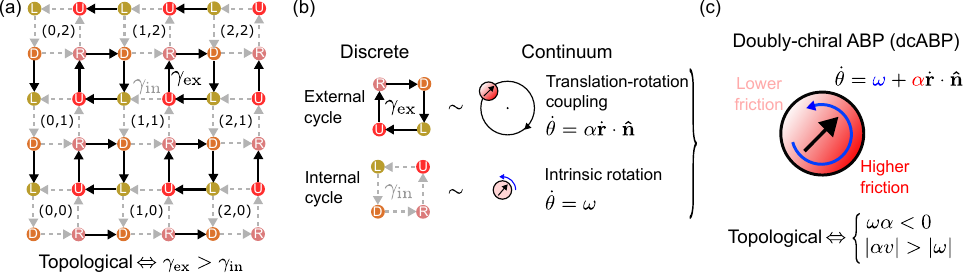}
	\caption{(a) Discrete lattice stochastic model introduced in Ref.~\citenum{Tang2021TopologyProtectsChiralCurrents}, which was shown to develop topologically protected edge currents in the regime $\gamma_\mathrm{ex}>\gamma_\mathrm{in}$. The internal states U,R,D,L stand for up, right, down, left. (b) External cycles in this model can be interpreted as causing simultaneous translation and rotation, and are thus analogous to translation-rotation coupling in the continuum limit. Internal cycles in the discrete model cause rotation without translation, and are thus analogous to intrinsic rotation in the continuum limit. (c) A doubly chiral ABP (dcABP) includes both sources of rotation, where translation-rotation coupling can arise from asymmetric friction. Analogy to the discrete model suggests that topologically protected edge currents may arise when $\omega\alpha<0$ (opposite chirality of external and internal cycles) and $|\alpha v| > |\omega|$ ($\gamma_\mathrm{ex}>\gamma_\mathrm{in}$).
	}\label{fig:fig3}
\end{figure*}

\subsection{Topologically protected edge currents in discrete stochastic systems}

Topologically protected edge currents, robust to boundary deformations, have been previously found in a particular model of discrete-space (lattice) classical stochastic dynamics out of equilibrium \cite{Tang2021TopologyProtectsChiralCurrents}. The corresponding lattice is depicted in Fig.~\ref{fig:fig3}(a). It consists of a four-state unit cell (with the four internal states here labeled U, D, L, R for up, down, left, right) repeated in a two-dimensional square lattice. There are two types of transitions, labeled external (with rate $\gamma_\mathrm{ex}$, which move across unit cells in the direction indicated by the current internal state, e.g.~U moves up, D moves down, and so on) and internal (with rate $\gamma_\mathrm{in}$, which remain within a unit cell), which correspondingly form external and internal cycles. Two key requirements where found to be needed for topologically protected edge currents to exist. First, external and internal cycles have opposite chiralities: e.g., if the external cycle is clockwise (U$\rightarrow$R$\rightarrow$D$\rightarrow$L$\rightarrow$U) as in Fig.~\ref{fig:fig3}(a), then the internal cycle must be counterclockwise (U$\rightarrow$L$\rightarrow$D$\rightarrow$R$\rightarrow$U) for an edge current to exist. Second, the internal transitions must be slower than the external ones, with $\gamma_\mathrm{ex}>\gamma_\mathrm{in}$.

The above considerations, when transferred to the continuum dynamics of an active particle, suggest that, for topologically protected edge currents to exist [Fig.~\ref{fig:fig3}(b,c)]: (i) the particle needs to have two sources of chirality, one of which corresponds to intrinsic rotation even in the absence of translation (internal cycles), the other to a translation-rotation coupling that causes rotation only concomitantly with translation (external cycles); (ii)  these two sources of chirality must have opposite sign; and (iii) the intrinsic chirality must be weaker than that arising from translation-rotation coupling. In the rest of this article, we construct and examine the dynamics of such a doubly chiral active particle.

\section{Doubly Chiral Active Brownian Particles \label{sec:dcabps}}

\subsection{Model}

As anticipated, we consider chiral ABPs that not only have an intrinsic angular velocity but also a translation-rotation coupling that results in cross-alignment of the particle to its instantaneous velocity. Such a cross-alignment term can result from substrate friction that is not symmetric relative to the self-propulsion axis, see Fig.~\ref{fig:fig3}(c) for an illustration and Section~\ref{sec:friction} and Appendix~\ref{app:friction} for a derivation. Because of the two sources of chirality, we term them doubly chiral ABPs (dcABPs). The overdamped dynamics of the particle is now defined by
\begin{eqnarray}
    \dot{\rr} & = & v \nn + \mu \mathbf{F} \label{eq:rdyn3} \\
    \dot{\theta} & = & \omega + \alpha \dot{\rr} \cdot \nn + \sqrt{2D_r} \eta  \label{eq:thetadyn} \\
    & = & \omega + \alpha v + \alpha \mu \mathbf{F} \cdot \nn  +  \sqrt{2\Dr} \eta \nonumber
\end{eqnarray}
where the only difference with respect to (\ref{eq:rdyn}--\ref{eq:thetadyn_cABP}) is the addition of the term with the parameter $\alpha$, which is the strength of cross-alignment to the instantaneous velocity  and provides the second source of chirality. We note that, away from boundaries or other particles, the undisturbed particle rotates with angular velocity $\omega_\mathrm{tot} \equiv \omega + \alpha v$, so that it describes bulk orbits with radius $R_\mathrm{bo} \equiv v/|\omega_\mathrm{tot}|$.

\subsection{Sliding mode at a straight boundary}

We now consider the deterministic dynamics ($D_r=0$) of a dcABP near a straight boundary, which without loss of generality we consider to be located at $x=0$ and keeping the particle to its right. The boundary induces a force $\mathbf{F}=(F_x(x),0)$ such that $F_x(x)$ is continuous and monotonically decreasing, and satisfying $F_x(x>0)=0$. The equations of motion (\ref{eq:rdyn3}-\ref{eq:thetadyn}) can be rewritten as:
\begin{eqnarray}
    \dot{x} & = & v \cos\theta + \mu F_x(x) \label{eq:xdynwall} \\
    \dot{y} & = & v \sin\theta  \label{eq:ydynwall} \\
    \dot{\theta} & = & \omega + \alpha (v + \mu F_x(x) \cos\theta )  \label{eq:thetadynwall}
\end{eqnarray}

Note that the dynamics of $y$ depends on $x$ and $\theta$, but not the other way around, and thus the dynamics of $y$ play only a passive role. We look for fixed points $(x_*,\theta_*)$ in the dynamics of $x$ and $\theta$, which implies
\begin{eqnarray}
    0 & = & v \cos\theta_* + \mu F_x(x_*) \label{eq:xfpwall} \\
    0 & = & \omega + \alpha (v + \mu F_x(x_*) \cos\theta_* )  \label{eq:thetafpwall}
\end{eqnarray}
Because $F_x(x)$ grows monotonically away from 0 as $x$ becomes negative, there will always be a value $x_*<0$ that solves (\ref{eq:xfpwall}), with $\theta_* \in (\pi/2,3\pi/2)$ so that $\cos \theta_*<0$. Substituting (\ref{eq:xfpwall}) into (\ref{eq:thetafpwall}), and rearranging, we obtain
\begin{equation}
    \sin \theta_* = \pm \sqrt{- \frac{\omega}{\alpha v}} \label{eq:fixedpoint}
\end{equation}
which shows that, for a pair of fixed points to exist, we need that $\omega$ and $\alpha$ have opposite signs ($\omega\alpha<0$, otherwise there is no solution as the r.h.s.~becomes imaginary) and moreover that $|\alpha v| > |\omega|$ (otherwise there is no solution as the r.h.s.~becomes $>1$ or  $<-1$). Note that the latter condition implies that the sign of $\alpha$ must control the sign of $\omega_\mathrm{tot}$, i.e.~the chirality of the cyclotron orbits in the bulk is determined by translation-rotation coupling. These conditions confirm precisely the expectations derived from the analogy to topologically protected edge currents in discrete systems [Fig.~\ref{fig:fig3}(c)].

The two solutions in (\ref{eq:fixedpoint}) correspond to a stable fixed point and a saddle point, as can be understood by studying the Jacobian matrix of the dynamical system evaluated at the fixed point, whose determinant is $\det(J_*) = 2 \mu |\cos \theta_*||F_x'|v \alpha \sin\theta_*$. The sign of the determinant is governed by the sign of the product $\alpha \sin\theta_*$. In the case of counterclockwise orbits in the bulk ($\omega_\mathrm{tot}>0$, and thus $\alpha>0$ if there exists a fixed point), then the solution with $\sin \theta_* >0$, or more precisely $\theta_* \in (\pi/2,\pi)$, is the stable solution. At this stable fixed point (sliding mode), the particle moves upwards along the left boundary with a speed
\begin{equation}
    v_y = v |\sin\theta_*| = \sqrt{-\frac{v\omega}{\alpha}}
\end{equation}
and would conversely move downwards with the same speed along a right boundary, with $\theta_* \in (-\pi/2,0)$. Thus, the sliding modes have opposite chirality (clockwise) to the counterclockwise bulk orbits.

In the limiting case where $\omega \to 0$, the two fixed points merge at $\theta_*=\pi$ (particle pointing directly against the boundary) and annihilate. In the other limiting case where $\omega \to -\alpha v$, the fixed points go to $\theta_*=\pi/2$ and $3\pi/2$ which, through (\ref{eq:xfpwall}), corresponds to $F_x(x_*)=0$ and thus $x_*=0$, i.e.~the fixed points move ``out'' of the boundary potential and disappear. For a harmonic potential with $F_x(x<0)=-kx$, we can directly solve for $x_*$ in (\ref{eq:xfpwall})  to obtain $x_* = -\frac{v}{\mu k} \sqrt{1+\frac{\omega}{\alpha v}}$. The particle penetrates deepest into the boundary to $x_*=-\frac{v}{\mu k}$ when $\omega \to 0$, and as already mentioned no longer penetrates the boundary ($x_* \to 0$) when $\omega \to -\alpha v$.

\begin{figure*}
	\centering
	\includegraphics[width=1\textwidth]{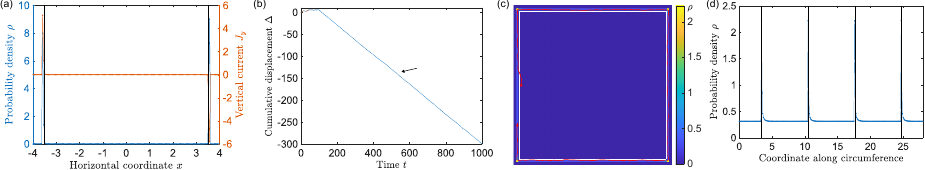}
	\caption{Robust topologically protected edge currents in dcABPs. (a) Steady state probability density (blue) and vertical current (red) along the horizontal coordinate in channel geometry, showing strong accumulation and local currents with opposite chirality to bulk orbits at the boundaries. (b) Evolution of the signed cumulative displacement in the vertical direction over time in the same simulation, which shows net drift in the positive direction when the particle is within one bulk orbit radius from the channel centerline (red segments) and in the negative direction when the particle enters a long-lived transport-inducing sliding mode along the boundaries (black arrows). (c) Steady state probability density in square confinement. One example of a particle trajectory, starting from the middle of the left boundary, is overlaid, and shows clockwise boundary-sliding with no backscattering at corners. (d) Steady state probability density along the circumference of the system (just outside the boundary, in clockwise direction), with no signatures of backscattering at corners (vertical black lines). For parameter values used in simulations, see Appendix~\ref{app:simulationdetails}.
	}\label{fig:fig4}
\end{figure*}

\subsection{Stochastic dynamics}

Results for the stochastic dynamics in the presence of noise ($D_r>0$) in channel and square geometries are shown in Fig.~\ref{fig:fig4}, and can be directly compared to those for chiral active rods (self-aligning cABPs) in Fig.~\ref{fig:fig2}. In a channel geometry, we find significantly stronger localization and currents near the boundaries (Fig.~\ref{fig:fig4}(a), note the values of the y-axes) and much more persistent sliding modes [Fig.~\ref{fig:fig4}(b)]. In addition, in a square geometry we find that there is no longer any backscattering at inside corners (Fig.~\ref{fig:fig4}(c,d) and Movie S3). Indeed, dcABPs in the sliding mode can easily turn both inside and outside corners without backscattering, implying topological protection. This allows them, for instance, to solve a simply connected maze by closely following its boundary (Movie S4).

\begin{figure}
	\centering
	\includegraphics[width=1\linewidth]{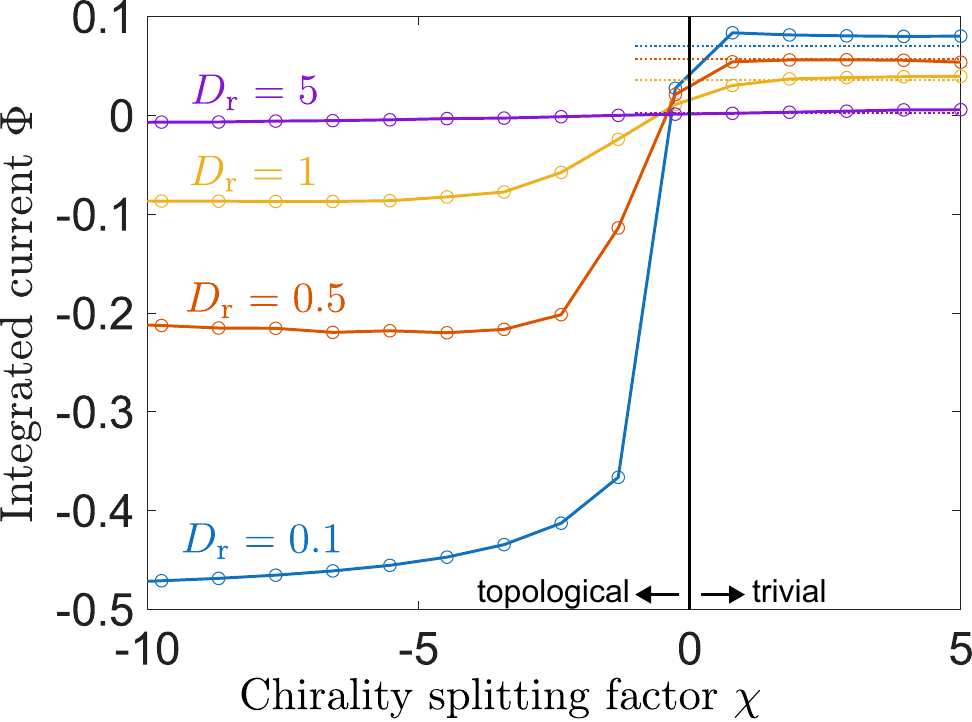}
	\caption{Integrated current $\Phi$ obtained in stochastic simulations in a channel geometry, as a dcABP crosses from the trivial to the topological regime, for several values of rotational noise. The total angular velocity of the particle in the bulk is fixed to $\omega_\mathrm{tot}=1$, so that the bulk orbit radius is constant throughout the transition, while we set $\omega=\omega_\mathrm{tot}\chi$ and $\alpha v = \omega_\mathrm{tot}(1-\chi)$, where $\chi$ is a chirality splitting factor. In the trivial regime (no sliding mode, $\chi\geq0$), $\Phi$ plateaus to the `spurious' (non-transport-related) value given by Eq.~(\ref{eq:intcurrent}) (dotted lines). In the topological regime (with sliding mode, $\chi<0$), $\Phi$ develops opposite chirality to bulk orbits and reflects true transport. For parameter values used in simulations, see Appendix~\ref{app:simulationdetails}.
	}\label{fig:fig5}
\end{figure}

To characterize how edge currents emerge as the system moves from the trivial (no sliding mode) to the topological (with sliding mode) phase, we measure the integrated current $\Phi$ in a channel geometry and with fixed speed $v$ and total angular velocity $\omega_\mathrm{tot}=\omega + \alpha v$ (and thus fixed bulk orbit radius $R_\mathrm{bo}$), as a function of a chirality splitting factor $\chi$ that controls the strength of intrinsic rotation vs translation-rotation coupling, such that $\omega=\omega_\mathrm{tot}\chi$ and $\alpha v = \omega_\mathrm{tot} (1-\chi)$. In this parametrization, the topological regime corresponds to $\chi<0$, as then we find both $\omega \alpha<0$ and $|\alpha v|>|\omega|$. We note as well that simple cABPs correspond to $\chi=1$ in this notation. The results are shown in Fig.~\ref{fig:fig5}, for several values of the noise strength $D_r$. We find that, in the  trivial regime $\chi>0$, the integrated edge current $\Phi$ rapidly plateaus to the `spurious' value (positive, i.e.~same chirality as bulk orbits) corresponding to bulk orbits at the center of the channel, given by Eq.~(\ref{eq:intcurrent}). As the system enters the topological regime $\chi<0$, $\Phi$ rapidly changes sign (negative, i.e.~opposite chirality to bulk orbits) and reaches a new plateau, the value of which depends on $D_r$. For small noise ($D_r \to 0$), this plateau value approaches $\Phi = -0.5$ (in dimensionless units with $\omega_\mathrm{tot}=v=1$), which is what is expected if the probability density is fully localized at the boundaries (half on each channel boundary) and the sliding speed along the boundary approaches the self-propulsion speed $v$.

\subsection{Sliding mode at curved boundaries}

In order to better understand how the sliding mode behaves in boundaries with more complex shapes, we carry out the fixed point calculation for a circular boundary with radius $R$, which could represent a particle in circular confinement or a particle interacting from the outside with a circular obstacle.

In polar coordinates with radial coordinate $r$, polar coordinate $\phi$, and particle orientation relative to the radial direction $\delta = \theta - \phi$, and in the absence of noise ($D_r=0$), the equations of motion (\ref{eq:rdyn3}-\ref{eq:thetadyn})  become
\begin{eqnarray}
    \dot{r} & = & v \cos\delta + \mu F_r(r) \label{eq:xdyncwall} \\
    r\dot{\phi} & = & v \sin\delta  \label{eq:ydyncwall} \\
    \dot{\delta} & = & \omega + \alpha (v + \mu F_r(r) \cos\delta ) - \frac{v\sin\delta}{r}  \label{eq:thetadyncwall}
\end{eqnarray}
where again we see that the polar coordinate $\phi$ plays only a passive role. Looking for a fixed point in the $(r,\delta)$ dynamics, and assuming a hard boundary with radius $R$ so that the fixed point occurs at $r_*\approx R$, we find that the orientational fixed points are given by
\begin{equation}
    \sin \delta_* = \frac{1}{2\alpha R} \pm \sqrt{\frac{1}{(2\alpha R)^2} - \frac{\omega}{\alpha v}}. \label{eq:fixedpointc}
\end{equation}
Two distinct cases must be considered: a particle inside circular confinement (i.e.~$F_r(r<R)=0$ and $F_r(r>R)<0$), and a particle interacting from the outside with a circular obstacle (i.e.~$F_r(r<R)>0$ and $F_r(r>R)=0$). For the inside case, the negative and positive solutions in (\ref{eq:fixedpointc}) correspond to the stable fixed point and the saddle point, respectively, while the opposite is true for the outside case.

\begin{figure*}
	\centering
	\includegraphics[width=1\textwidth]{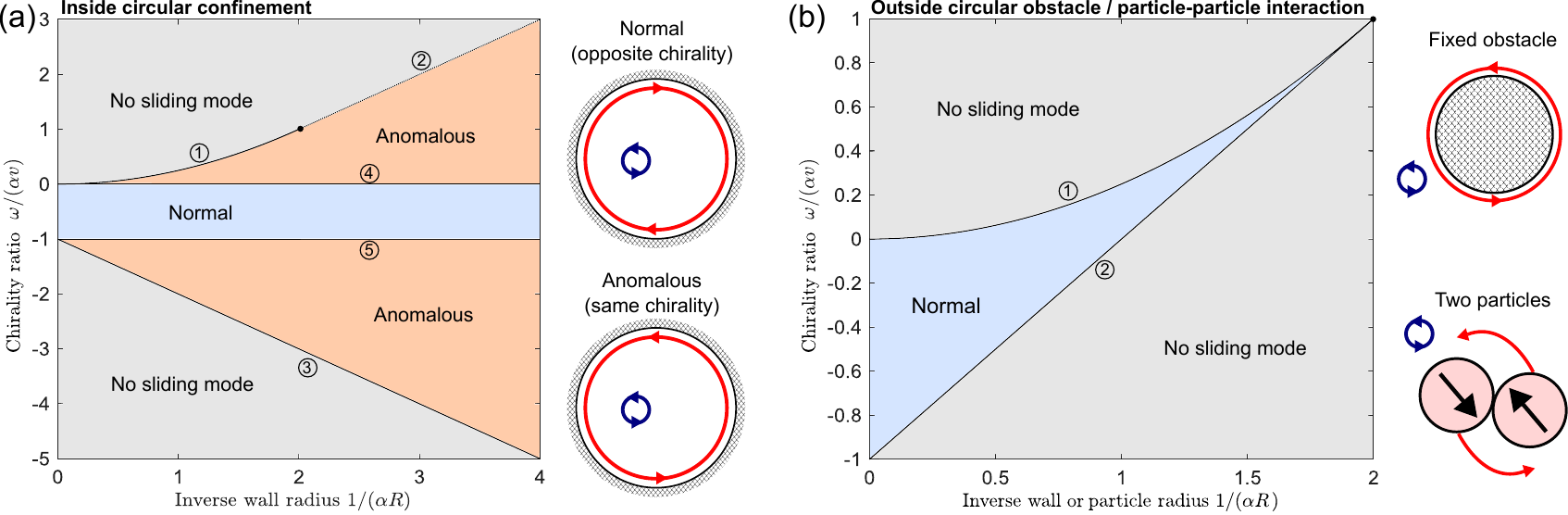}
	\caption{Sliding modes along curved boundaries. (a) Phase diagram for the existence and properties of a deterministic sliding mode for a dcABP inside circular confinement. Normal or anomalous sliding modes correspond to the sliding mode having opposite or equal chirality to that of bulk orbits, respectively.  (b) Phase diagram that describes both the case of a dcABP interacting with a fixed circular obstacle (of radius $R$) or of two dcABPs (of radius $R$) interacting with each other. In the obstacle case, the sliding mode has equal chirality to the bulk orbits. In the two-particle case, the sliding mode corresponds to a spinning mode, where the pair spins with the same chirality as the bulk orbits. In both panels, the circled numbers refer to the boundary lines defined in (\ref{eq:circles1}--\ref{eq:circles2}).
	}\label{fig:fig6}
\end{figure*}

Investigating the limits in which the two fixed points in (\ref{eq:fixedpointc}) are well defined, we derive the boundary lines
\begin{eqnarray}
\frac{\omega}{\alpha v} \overset{\Circled{1}}{=} \frac{1}{(2\alpha R)^2},~~\frac{\omega}{\alpha v} \overset{\Circled{2}}{=} -1 + \frac{1}{\alpha R},~~\frac{\omega}{\alpha v} \overset{\Circled{3}}{=} -1 - \frac{1}{\alpha R}. \label{eq:circles1}
\end{eqnarray}
Additionally, by considering the chirality of the edge currents and bulk orbits, we derive
\begin{eqnarray}
\frac{\omega}{\alpha v} \overset{\Circled{4}}{=} 0,~~\frac{\omega}{\alpha v} \overset{\Circled{5}}{=} -1.  \label{eq:circles2}
\end{eqnarray}
These five boundary lines can be used to define a phase diagram for the existence of a sliding mode.

We take $\alpha>0$ in the following without loss of generality. For the case of a particle inside circular confinement, we find the phase diagram in Fig.~\ref{fig:fig6}(a), defined by all five boundary lines. For $0>\frac{\omega}{\alpha v}>-1$, which corresponds to the regime of existence of a sliding mode along a straight boundary, a normal sliding mode (with chirality opposite to that of the bulk) exists independently of the curvature of the boundary, which can be arbitrarily large [blue region in Fig.~\ref{fig:fig6}(a)]. This serves as additional analytical confirmation that dcABPs can turn along arbitrarily sharp inside corners. Interestingly, the range of $\frac{\omega}{\alpha v}$ for which a sliding mode exists broadens as the boundaries become more curved, but this corresponds to an anomalous sliding mode with chirality equal to that of the bulk [orange regions in Fig.~\ref{fig:fig6}(a)]. Note that, as the boundary $\Circled{4}$ is crossed, the chirality of the sliding mode flips while that of the bulk orbits remains the same, whereas the opposite is true as the boundary $\Circled{5}$ is crossed. Simulations showing normal and anomalous sliding modes are shown in Movies S5 and S6, respectively.

For the case of a particle outside a circular obstacle, we find the phase diagram in Fig.~\ref{fig:fig6}(b), defined only by the boundary lines  $\Circled{1}$ and $\Circled{2}$. Only normal sliding modes exist (which in this case correspond to sliding modes with the same chirality as the bulk orbits), in a significantly reduced region of parameter space relative to the case of inside confinement. A simulation showing such a sliding mode is shown in Movie S7. Most notably, no sliding mode exists beyond boundary line $\Circled{2}$, which is natural as this line corresponds to the parameter values for which the radius of the bulk orbits equals the radius of the obstacle, $R_\mathrm{bo}=R$. Beyond this line, the circular obstacle is smaller than the bulk orbits, and no sliding mode exists. This effect has consequences for the behavior of dcABPs when they turn sharp outside corners, as it implies that they cannot turn such corners while remaining in contact with the boundary at all times. Indeed, simulations show that, at sharp outside corners, dcABPs briefly leave the boundary but immediately circle back towards it (describing a segment of a bulk orbit).

Note that, throughout this section, we have assumed that the particles are point-like. To account for a finite particle radius, the effective radius of the circular boundary must be decreased or increased by one particle radius, respectively for the inside and outside cases.

\subsection{Interparticle interactions}

Because dcABPs become bound to and slide along circular obstacles, it is interesting to consider whether this can lead to bound states when two dcABPs come into contact. To this end, we look for fixed points in the relative distance between the particles $\ell$ and the orientations $\delta_1$ and $\delta_2$ of the two particles relative to the line joining their centers. In this case, both the position of the center of mass $\rr_\mathrm{com}=(\rr_1+\rr_2)/2$ of the two particles and the orientation $\phi$ of the center-to-center line relative to the lab frame play only a passive role. Upon a change of coordinates in (\ref{eq:rdyn3}-\ref{eq:thetadyn}), the dynamics of these relative coordinates are given by
\begin{eqnarray}
    \dot{\rr}_\mathrm{com} =   v (\nn_1+\nn_2) \label{eq:rcomdyn2} \\
     \ell \dot{\phi}  =  - v (\sin\delta_1 + \sin\delta_2)  \label{eq:phidyn2} \\
     \dot{\ell}  =  - v (\cos\delta_1 + \cos\delta_2) + 2 \mu F_\ell(\ell) \label{eq:elldyn2} \\
     \dot{\delta}_i =  \omega + \alpha (v - \mu F_\ell(\ell) \cos\delta_i ) + \frac{v}{\ell}(\sin\delta_1+\sin\delta_2)  \label{eq:delta1} 
\end{eqnarray}

We note that $\dot{\delta}_1=\dot{\delta}_2=0$ can only be satisfied if $\cos\delta_1 = \cos\delta_2$, which implies either $\delta_1=\delta_2$ or $\delta_1=-\delta_2$. We find that only the former case leads to a stable fixed point, while the latter only results in saddle points. Setting $\delta=\delta_1=\delta_2$, and assuming hard particles so that the fixed point in their separation satisfies $\ell_*=2R$ with $R$ the particle radius, we find that there exist fixed points at the relative orientations
\begin{equation}
    \sin \delta_* = - \frac{1}{2\alpha R} \pm \sqrt{\frac{1}{(2\alpha R)^2} - \frac{\omega}{\alpha v}}. \label{eq:fixedpoint2}
\end{equation}
which we note coincides up to a change of sign with (\ref{eq:fixedpointc}), although $R$ now represents the particle radius, rather than the boundary radius. Without loss of generality, let us focus on the case with $\alpha>0$. In this case, it is the solution with a negative sign that corresponds to the stable fixed point, which now corresponds to both particles spinning around each other (spinning mode). Analyzing the conditions for its existence, we recover an identical phase diagram to that for a single particle outside a circular obstacle, see Fig.~\ref{fig:fig6}(b). A simulation showing the spinning mode is shown in Movie S8. In the spinning mode, the pair spins with the same chirality as the bulk orbits, while their center of mass remains static.

\section{Practical implementation of Doubly Chiral Active Particles \label{sec:practical}}

\subsection{Detailed mechanical model \label{sec:friction}}

Having explored in detail the phenomenology of the dcABP model as given by equations (\ref{eq:rdyn3}-\ref{eq:thetadyn}), we now turn to the question of how such dynamics can arise in practice for an active particle. We have anticipated that a translation-rotation coupling such as the one that $\alpha$ describes can arise as a consequence of a left-right asymmetry in the friction of the particle (e.g.~a friction gradient in the direction perpendicular to the axis of propulsion of the particle), see Fig.~\ref{fig:fig3}(c).

Indeed, in Appendix~\ref{app:friction}, we derive the full friction tensor for an active particle with an arbitrary friction distribution. For a friction density $\tilde{\xi}(\xx)$ where $\xx$ is a body-fixed coordinate, we can calculate the total translational friction $\xi$ as $\xi = \int \mathrm{d}\xx\, \tilde{\xi}(\xx)$, the total rotational friction $\xi_r$ as $\xi_r = \int \mathrm{d}\xx\, \tilde{\xi}(\xx) \xx^2$, and the location of the \emph{center of friction} $\mathbf{a}$ relative to the geometric center as
\begin{equation}
    \mathbf{a} = a_\parallel \nn + a_\perp \nn_\perp = \frac{1}{\xi} \int \mathrm{d}\xx\, \tilde{\xi}(\xx) \xx \label{eq:cof}
\end{equation}
where, as before, $\nn = (\cos \theta,\sin \theta)^t$ is the propulsion direction, and we introduce the unit vector perpendicular to it $\nn_\perp = (-\sin \theta,\cos \theta)^t$. Self-aligning active particles correspond to the case with $a_\perp=0$ and $a_\parallel<0$, where the center of friction lies along the axis of propulsion but trails behind the geometric center. Self-anti-aligning active particles, in turn, correspond to $a_\perp=0$ and $a_\parallel>0$, where the center of friction lies along the axis of propulsion but is ahead of the geometric center. The case of interest for our purpose is that of cross-aligning active particles, which corresponds to $a_\parallel=0$ and $a_\perp \neq 0$, so that the center of friction lies on the line perpendicular to the axis of propulsion that passes through the center of the particle.

In this particular case, with $a_\parallel=0$ and $a_\perp \neq 0$, and for circular particles for which external forces produce no torque about the geometric center, the deterministic equations of motion for the active particle become
\begin{equation}
\begin{pmatrix}
\mathbf{F}+f\nn \\
\tau 
\end{pmatrix}
=
\begin{pmatrix}
\xi \mathbf{1} & -\xi a_\perp \nn \\
-\xi a_\perp \nn^t & \xi_r 
\end{pmatrix}
\begin{pmatrix}
\dot{\rr} \\
\dot{\theta}
\end{pmatrix}
\label{eq:eomfriction}
\end{equation}
where $\mathbf{1}$ is the identity matrix of order 2, $f>0$ is the net self-propulsion force (directed along $\nn$ by definition of $\nn$), $\tau$ is the net self-exerted torque about the geometric center, and as before $\mathbf{F}$ is the external force. We note that the existence of a well-defined friction tensor enables the addition of thermodynamically consistent noise via the fluctuation-dissipation theorem, but we do not pursue this direction here.

Let us consider again the existence of a sliding mode along a straight boundary. Expressing $\rr$ in its components $(x,y)$ and setting $\dot{x}=\dot{\theta}=0$ in (\ref{eq:eomfriction}), we immediately obtain
\begin{equation}
    \sin \theta_* = \pm \sqrt{-\frac{\tau}{a_\perp f}} \label{eq:fixedpointfriction}
\end{equation}
which is found to be exactly analogous to the result (\ref{eq:fixedpoint}) obtained for the dcABP model. Indeed, (\ref{eq:fixedpointfriction}) implies that, for a sliding mode to exist, we need that the intrinsic torque ($\tau$) and the torque due to translation-rotation coupling ($a_\perp f$) have opposite signs, and moreover that the intrinsic torque be the weaker of the two ($|\tau|<|a_\perp f|$).

To further pin down the correspondence between the mechanically detailed model (\ref{eq:eomfriction}) and the phenomenological dcABP model (\ref{eq:rdyn3}-\ref{eq:thetadyn}), we rewrite (\ref{eq:eomfriction}) in mobility form, inverting the friction tensor to obtain
\begin{equation}
\begin{pmatrix}
\dot{\rr} \\
\dot{\theta}
\end{pmatrix}
=
\frac{1}{\Delta}
\begin{pmatrix}
\xi_r \mathbf{1} - \xi a_\perp^2(\mathbf{1}-\nn\nn) & \xi a_\perp \nn \\
\xi a_\perp \nn^t & \xi 
\end{pmatrix}
\begin{pmatrix}
\mathbf{F}+f\nn \\
\tau 
\end{pmatrix}
\label{eq:eommobility}
\end{equation}
where $\Delta \equiv \xi\xi_r - \xi^2 a_\perp^2>0$ is the determinant of the friction tensor (positive due to its positive definiteness).

By defining the translational velocity $v \equiv f/\xi$, the angular velocity $\omega \equiv \tau / \xi_r$, the translational mobility $\mu \equiv 1/\xi$, the translation-rotation coupling $\alpha \equiv \xi a_\perp / \xi_r$, and the dimensionless coupling strength $\Gamma \equiv \xi a_\perp^2 / \xi_r$, we can rewrite (\ref{eq:eommobility}) as
\begin{eqnarray}
    \dot{\rr} & = & \frac{1+\Gamma \frac{\omega}{\alpha v}}{1-\Gamma} v \nn + \frac{\mathbf{1}-\Gamma(\mathbf{1}-\nn\nn)}{1-\Gamma}\mu \mathbf{F} \label{eq:rdyn3b} \\
    \dot{\theta} & = & \frac{1}{1-\Gamma}\omega + \frac{1}{1-\Gamma} \alpha(v + \mu \mathbf{F} \cdot \nn)  \label{eq:thetadynb}
\end{eqnarray}
where we note that we always have $0 \leq \Gamma < 1$ due to the positive definiteness of the friction tensor. As can be directly noticed by inspection, the mechanically detailed equations (\ref{eq:rdyn3b}--\ref{eq:thetadynb}) become identical to the phenomenological dcABP model (\ref{eq:rdyn3}-\ref{eq:thetadyn}) in the limit of small coupling $\Gamma \ll 1$ (provided that $\frac{\omega}{\alpha v}$ remains $O(1)$, which importantly includes the topological regime).

\begin{figure}
	\centering
	\includegraphics[width=1\linewidth]{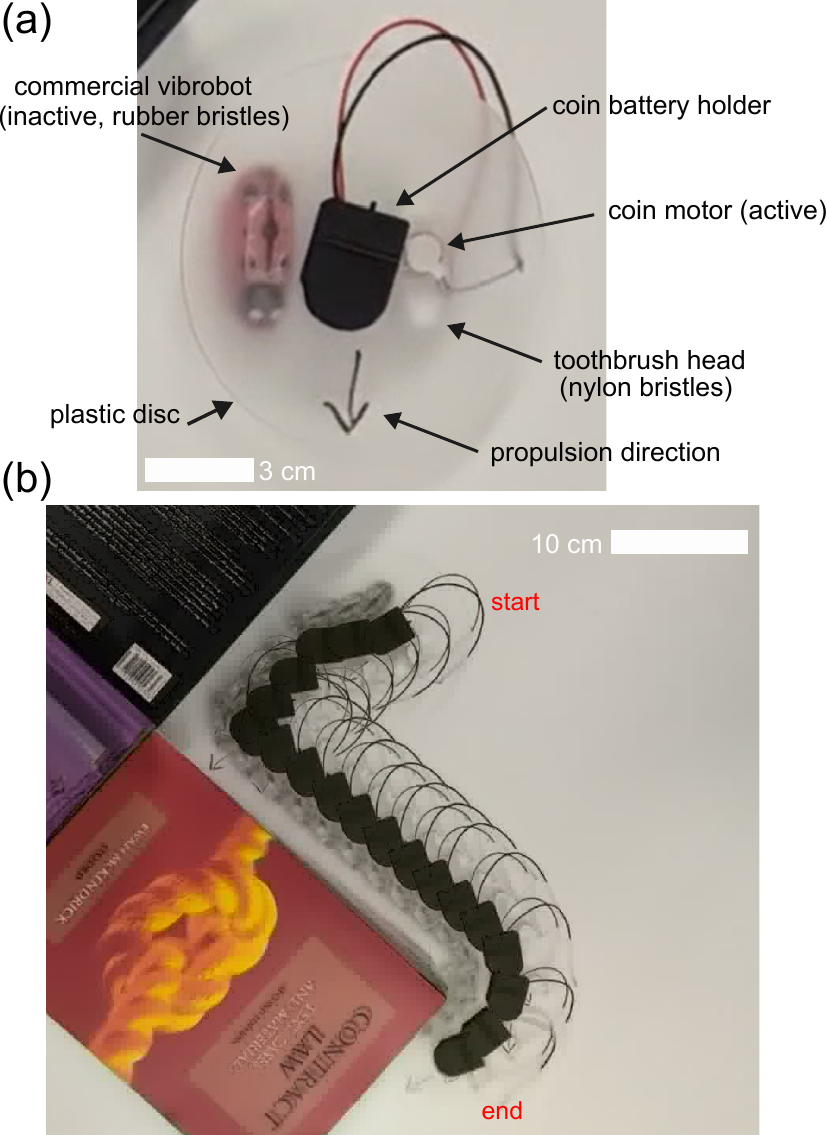}
	\caption{Experimental realization of a doubly chiral vibrobot. (a) Schematic describing a functioning design, see text for further details. (b) Time-lapse image ($\simeq 40$ s in real time) showing how the vibrobot slides along straight boundaries and can turn both inside and outside corners. See Movie S9 for the corresponding video.
	}\label{fig:fig7}
\end{figure}

\subsection{Experiment: Doubly chiral vibrobots}

The above mechanically detailed model suggests a blueprint towards building dcABPs in experiment. We chose to implement it by mounting vibrobots, both commercial (Kinberry Mini Robot Beetle Interactive Cat Toy) and made in-house (toothbrush heads with a coin vibrating motor operating at 3 V, 70 mA, 12000 rpm) onto a round plastic disc (10 cm diameter and 1 mm thickness), using double-sided sticky tape.

Within about 20 minutes of experimentation, we settled on a functioning design, which consisted of an \emph{active} (powered on) in-house vibrobot with nylon bristles and thus low friction, mounted to the left of the geometric center, and an \emph{inactive} (powered off) commercial vibrobot with rubber bristles and thus high friction, mounted to the right of the geometric center, see Fig.~\ref{fig:fig7}(a). Both were mounted on the line perpendicular to the propulsion direction that passes through the geometric center. The active in-house vibrobot provided both the propulsion and a counterclockwise intrinsic torque (importantly, single vibrobots typically show chirality, because of imperfections in their bristle arrangement \cite{274x-4vgn}) while the inactive commercial vibrobot simply provided friction, causing a translation-rotation coupling with strong clockwise chirality.

A time-lapse image of the resulting doubly chiral vibrobot sliding along an arrangement of books, turning both inside and outside corners, is shown in Fig.~\ref{fig:fig7}(b). A longer video of the same experiment is shown in Movie S9. Overall, we found that doubly chiral vibrobots are easy to construct and require very little fine tuning, as long as one ensures an asymmetric distribution of friction in the direction perpendicular to the propulsion direction.

\section{Conclusion}

We have introduced here doubly chiral active particles (dcABPs) as a new type of particle with a particularly robust tendency to undergo edge transport without backscattering, implying topological protection. This stands in contrast with simple cABPs, which we have shown to not undergo boundary-induced transport despite the existence of nonzero currents near boundaries in steady state, and chiral active rods or self-aligning cABPs, which we have shown to undergo boundary-induced transport but to backscatter at inside corners. 

We introduced dcABPs in analogy with discrete lattice out-of-equilibrium stochastic models that had previously been shown to exhibit topologically protected edge currents. Considering the existence of sliding modes in the deterministic dynamics of dcABPs, we showed that this analogy carries over not only qualitatively but even mathematically, with the conditions $\omega\alpha<0$ and $|\alpha v|>|\omega|$ for the existence of a boundary-sliding mode being directly related to the conditions for the existence of topological protection in the lattice model, see Fig.~\ref{fig:fig3}. Whether similar translations between the dynamics and phenomenology of other models of continuum active systems and out-of-equilibrium discrete systems, in either direction, constitutes an interesting avenue for future work.

In dcABPs, the first source of chirality corresponds to an intrinsic active rotation, whereas the second source of chirality corresponds to a translation-rotation coupling that causes cross-alignment to the instantaneous velocity. Far from constituting an ad-hoc set of dynamics, we showed that double chirality arises very naturally for active particles that have an asymmetric friction distribution in the direction perpendicular to the direction of motion. As such, future work may investigate whether double chirality can arise in other types of active particles that are effectively described by overdamped equations of motion, including e.g. self-diffusiophoretic active particles \cite{PhysRevE.82.015304}, or swimming  \cite{lauga2006swimming} and gliding \cite{lettermann2026chirality} cells. As we have shown, the manufacture of doubly chiral robots is uncomplicated and low-cost, and could thus be useful to the field of swarm robotics \cite{brambilla2013swarm}. A key practical advantage is that they can attach to and follow along boundaries without any need for proximity sensors, allowing for a swarm to e.g. quickly map the boundaries of any arbitrary location.

While this work was being prepared for submission, a preprint was made available online \cite{kuroda2026designingtopologicaledgecurrents} which independently introduces the dcABP dynamics (although they are interpreted as arising from a particle with time-dependent `switching' chirality) and demonstrates the existence of topologically protected edge currents under these dynamics. Interestingly, despite this common central point, the two works are entirely complementary. While we have focused here on how the dcABP dynamics arise mechanically from two competing sources of chirality (intrinsic rotation and asymmetric friction), how the edge currents arise as sliding modes in the deterministic limit, and how these dynamics lead to boundary-induced transport with no backscattering as opposed to those of standard cABPs and chiral active rods, Ref.~\citenum{kuroda2026designingtopologicaledgecurrents} instead focuses on the collective behavior of these particles and on the identification of the resulting edge currents as topological by analyzing the band structure in reciprocal space. We thus refer the interested reader to this work for a further exploration of the dynamics of dcABPs.

In summary, dcABPs constitute a promising avenue towards the development of active microparticles or swarm robots that can bind to and follow boundaries, thus allowing for the single-particle dynamics to turn from two-dimensional random exploration to one-dimensional directed motion. It remains to be seen whether existing microorganisms or cells make use of this strategy in their motion, or relatedly, whether more complex mechanisms for locomotion can result in dcABP-like effective dynamics at the coarse grained level, when only position and orientation are kept in the effective description.

\section*{Acknowledgements.}

We thank J.~Kocka, S.~Pritchard, S.~Redford, and V.~Traynor for participating in the vibrobot-building session. J.A.-C.~would like to thank the organisers and participants of the CECAM workshop ``Emergent dynamics of active colloids: Chirality, non-reciprocity and memory'' at CECAM-HQ-EPFL, Lausanne, Switzerland, for enabling interesting discussions and providing helpful comments on a presentation of this work.

\appendix

\section{\\Integrated current for cABPs with rotational noise \label{app:withnoise}}

We show that the full expression (with finite $\Dr$) for the integrated current $\Phi$ in Eq.~\ref{eq:intcurrent} can also be derived from a perspective based on bulk trajectories, without any reference to the existence of boundaries. We focus on the case of circular confinement, for which the calculation is most transparent.

In the deterministic limit, we showed above that the integrated flux was given by the number of orbits that enclose the origin per unit time, which was given by [see Eq.~\ref{eq:intcurrent_arena}]
\begin{equation}
    \Phi = \rho_\mathrm{b}\frac{A_\mathrm{bo}}{T}
\end{equation}
where $A_\mathrm{bo}$ is the area enclosed by a bulk orbit, and $T$ the period of a bulk orbit. In the presence of rotational noise, trajectories no longer close into orbits. However, we can simply replace $A_\mathrm{bo}/T$ by its stochastic analog, namely the mean signed area swept per unit time $\langle \dot{A}_\mathrm{bo} \rangle$, such that
\begin{equation}
    \Phi = \rho_\mathrm{b} \langle \dot{A}_\mathrm{bo} \rangle \label{eq:Phistoc}
\end{equation}
with
\begin{equation}
    \langle \dot{A}_\mathrm{bo} \rangle \equiv \frac{1}{2} \int_0^\infty \langle \dot{\rr}(0) \times \dot{\rr}(t) \rangle \cdot \zz  \, \mathrm{d}t.
\end{equation}
where the averages are with respect to noise realizations. This type of quantity has been previously used in the context of measuring entropy production in nonequilibrium trajectories \cite{gnesotto2020learning}.

For a cABP with dynamics given by (\ref{eq:rdyn}-\ref{eq:thetadyn_cABP}), and defining $\nn(\theta)=(\cos\theta,\sin\theta)^t$, we have
\begin{equation}
    \left[\dot{\rr}(0)\times \dot{\rr}(t)\right]\cdot\zz
    =
    v^2
    \left[\nn(\theta(0))\times \nn(\theta(t))\right]\cdot\zz .
\end{equation}
Using
\begin{equation}
    \left[\nn(\theta_1)\times \nn(\theta_2)\right]\cdot\zz
    =
    \sin(\theta_2-\theta_1),
\end{equation}
we obtain
\begin{equation}
    \left\langle
    \left[\dot{\rr}(0)\times \dot{\rr}(t)\right]\cdot\zz
    \right\rangle
    =
    v^2
    \left\langle
    \sin\left[\theta(t)-\theta(0)\right]
    \right\rangle .
\end{equation}
For the autonomous angular dynamics in (\ref{eq:thetadyn_cABP}), this average evaluates to
\begin{equation}
    \left\langle
    \sin\left[\theta(t)-\theta(0)\right]
    \right\rangle
    =
    e^{-\Dr t}\sin(\omega t).
\end{equation}
Therefore
\begin{equation}
    \left\langle \dot{A}_\mathrm{bo} \right\rangle
    =
    \frac{v^2}{2}
    \int_0^\infty
    e^{-\Dr t}\sin(\omega t)\,\mathrm{d} t,
\end{equation}
with the integral evaluating to
\begin{equation}
    \int_0^\infty
    e^{-\Dr t}\sin(\omega t)\,\mathrm{d} t
    =
    \frac{\omega}{\Dr^2+\omega^2}.
\end{equation}
The mean signed area swept per unit time is thus given by
\begin{equation}
    \left\langle \dot{A}_\mathrm{bo} \right\rangle
    =
    \frac{v^2\omega}{2(\Dr^2+\omega^2)}.
\end{equation}
Finally, substituting this into (\ref{eq:Phistoc}) we obtain
\begin{equation}
    \Phi
    =
    \frac{\rho_\mathrm{b}v^2\omega}
    {2(\omega^2+\Dr^2)} .
\end{equation}
which coincides with the full result in Eq.~\ref{eq:intcurrent}.

\section{\\Derivation of the friction tensor \label{app:friction}}

Let us consider a lab-fixed coordinate $\XX$ and a body-fixed coordinate $\xx = x_\parallel \nn + x_\perp \nn_\perp$, where $\nn = (\cos \theta,\sin \theta)^t$ is the propulsion direction (to be fixed a posteriori), and $\nn_\perp = (-\sin \theta,\cos \theta)^t$ a unit vector perpendicular to it. The two coordinates are related by $\XX=\rr + \xx$, where $\rr$ is the location of the geometric center of the particle in the lab frame. Taking the time derivative of this equation, we obtain
\begin{equation}
    \dot{\XX}=\dot{\rr} + (x_\parallel \nn_\perp - x_\perp \nn) \dot{\theta} \label{eq:geom}
\end{equation}
where we have used that $\dot{\nn}=\dot{\theta}\nn_\perp$ and $\dot{\nn}_\perp=-\dot{\theta}\nn$.

The body has friction density $\tilde{\xi}(\xx)$. The overdamped dynamics of any point of the particle are given by
\begin{equation}
    \tilde{\xi}(\xx)\dot{\XX}= \tilde{\mathbf{F}}(\xx) + \tilde{\mathbf{f}}(\xx) + \tilde{\mathbf{g}}(\xx) \label{eq:dyn}
\end{equation}
where $\tilde{\mathbf{F}}(\xx)$ is the external force density, $\tilde{\mathbf{f}}(\xx)$ the active propulsion force density, and $\tilde{\mathbf{g}}(\xx)$ are the constraint forces keeping the particle together.

Multiplying (\ref{eq:geom}) by $\tilde{\xi}(\xx)$ and integrating over the surface of the particle, we obtain
\begin{equation}
  \int \mathrm{d}\xx\, \tilde{\xi}(\xx) \dot{\XX}= \xi \dot{\rr} + \xi \dot{\theta} ( a_\parallel \nn_\perp - a_\perp \nn) \label{eq:app1}
\end{equation}
where $\xi = \int \mathrm{d}\xx\, \tilde{\xi}(\xx)$ is the total friction, and $a_\parallel$ and $a_\perp$ are the coordinates of the center of friction as defined in (\ref{eq:cof}). If, before integrating, we additionally take a cross product with $\xx$, we obtain
\begin{equation}
  \int \mathrm{d}\xx\, \tilde{\xi}(\xx) \xx \times \dot{\XX}= \xi \left[ (a_\parallel \nn_\perp - a_\perp \nn) \cdot \dot{\rr} \right] \zz + \xi_r \dot{\theta} \zz \label{eq:app2}
\end{equation}
where we have used the identities $\nn \times \dot{\rr} = (\nn_\perp \cdot \dot{\rr} )\zz$, $\nn_\perp \times \dot{\rr} = -(\nn \cdot \dot{\rr} )\zz$, $\nn \times \nn_\perp = \zz$, and $\nn_\perp \times \nn = -\zz$, and where $\xi_r = \int \mathrm{d}\xx\, \tilde{\xi}(\xx) \xx^2$ is the total rotational friction.

On the other hand, integrating  (\ref{eq:dyn}) over the surface of the particle, we obtain
\begin{equation}
  \int \mathrm{d}\xx\, \tilde{\xi}(\xx) \dot{\XX}= \mathbf{F} + f \nn \label{eq:app3}
\end{equation}
where $\mathbf{F}=\int \mathrm{d}\xx\, \tilde{\mathbf{F}}(\xx)$ is the total external force, $f\nn=\int \mathrm{d}\xx\, \tilde{\mathbf{f}}(\xx)$ is the total self-propulsion force (which serves to define $\nn$), and we have used that $\int \mathrm{d}\xx\, \tilde{\mathbf{g}}(\xx)=0$ as the constraint forces produce no net force. If, before integrating, we additionally take a cross product with $\xx$, we obtain
\begin{equation}
  \int \mathrm{d}\xx\, \tilde{\xi}(\xx) \xx \times \dot{\XX}= T \zz + \tau \zz \label{eq:app4}
\end{equation}
where $T\zz=\int \mathrm{d}\xx\, \xx \times \tilde{\mathbf{F}}(\xx)$ is the total external torque about the geometric center, $\tau \zz=\int \mathrm{d}\xx\, \xx \times \tilde{\mathbf{f}}(\xx)$ is the total active (self-exerted) torque about the geometric center, and we have used that $\int \mathrm{d}\xx \, \xx \times \tilde{\mathbf{g}}(\xx)=0$ as the constraint forces produce no net torque.

Equating (\ref{eq:app1}) to (\ref{eq:app3}) and (\ref{eq:app2}) to (\ref{eq:app4}), we finally obtain
\begin{equation}
\begin{pmatrix}
\mathbf{F}+f\nn \\
T+\tau 
\end{pmatrix}
=
\begin{pmatrix}
\xi \mathbf{1} & \xi (a_\parallel \nn_\perp - a_\perp \nn) \\
\xi(a_\parallel \nn_\perp - a_\perp \nn)^t & \xi_r 
\end{pmatrix}
\begin{pmatrix}
\dot{\rr} \\
\dot{\theta}
\end{pmatrix}
\end{equation}
where $\mathbf{1}$ is the identity matrix of order 2. Eq.~(\ref{eq:eomfriction}) in the main text corresponds to the special case $T=a_\parallel=0$ of this equation.

\section{\\Simulation details \label{app:simulationdetails}}

All simulations were performed using an Euler-Maruyama scheme, where $\mathrm{d}t$ denotes the numerical time step and $n_\mathrm{steps}$ the total number of steps in the simulation. The repulsive interactions with boundaries and between particles were implemented using harmonic repulsion, with spring constant $k$. In all Figures, we set $v=\omega_\mathrm{tot}=1$, which sets the bulk orbit radius $R_\mathrm{bo}=v/|\omega_\mathrm{tot}|=1$ as unit length scale and the inverse total angular velocity $\omega_\mathrm{tot}^{-1}=1$ as unit time scale. Note that, for simple cABPs and chiral rods, $\omega_\mathrm{tot}=\omega$, while for dcABPs $\omega_\mathrm{tot}=\omega + \alpha v$. The spring constant is set to $k=10$ in all cases. In channel geometry, $L$ denotes the width of the channel. In square geometry, $L$ denotes the side length of the square. In circular arena geometry, $R$ denotes the radius of the arena. Whenever binning was required for data analysis, the width or side length of the bin is denoted as $d_\mathrm{bin}$.
 
In Fig.~\ref{fig:fig1}(b,c), we use $L=7$, $\Dr=0.1$, $\mathrm{d}t=0.1$, $n_\mathrm{steps}=10^8$, and $d_\mathrm{bin}=0.05$ in (b). In Fig.~\ref{fig:fig1}(f,g), we use $R=4$, $\Dr=0.1$, $\mathrm{d}t=0.1$, $n_\mathrm{steps}=10^8$, and $d_\mathrm{bin}=0.05$ in (f).

In Fig.~\ref{fig:fig2}(a,b), we use $\zeta=10$, $L=7$, $\Dr=0.1$, $\mathrm{d}t=0.1$, $n_\mathrm{steps}=10^8$, and $d_\mathrm{bin}=0.05$ in (a). In Fig.~\ref{fig:fig2}(c,d), we use $\zeta=10$, $L=7$, $\Dr=0.1$, $\mathrm{d}t=0.1$, $n_\mathrm{steps}=10^8$, and $d_\mathrm{bin}=0.1$.

In Fig.~\ref{fig:fig4}(a,b), we use $\alpha=2$, $\omega=-1$, $L=7$, $\Dr=0.1$, $\mathrm{d}t=0.1$, $n_\mathrm{steps}=10^8$, and  $d_\mathrm{bin}=0.05$ in (a). In Fig.~\ref{fig:fig4}(c,d), we use $\alpha=2$, $\omega=-1$, $L=7$, $\Dr=0.1$, $\mathrm{d}t=0.1$, $n_\mathrm{steps}=10^8$, and $d_\mathrm{bin}=0.1$.

In Fig.~\ref{fig:fig5}, we use $L=7$, $\alpha=1-\chi$, $\omega=\chi$, $\mathrm{d}t=0.1$, and $n_\mathrm{steps}=10^8$.

\section{\\Description of the movies \label{app:moviedescription}}

\begin{itemize}
    \item Movie S1: Simulation of a simple cABP showing lack of boundary-induced transport. Parameters used: $\omega=-1$, $\Dr=0.01$, and $\mathrm{d}t=0.01$.
    \item Movie S2: Simulation of a self-aligning cABP (or chiral rod) showing boundary-induced transport along a straight boundary but backscattering at an inside corner. Parameters used: $\omega=-1$, $\zeta=5$, $\Dr=0.01$, and $\mathrm{d}t=0.01$.
    \item Movie S3: Simulation of a dcABP showing boundary-induced transport along a straight boundary and no backscattering at an inside corner. Parameters used: $\omega=1$, $\alpha=-2$, $\Dr=0.01$, and $\mathrm{d}t=0.01$. 
    \item Movie S4: Simulation of a dcABP solving a simply connected maze by following a boundary. Parameters used: $\omega=1$, $\alpha=-2$, $\Dr=0.1$, and $\mathrm{d}t=0.01$. The side length of the maze is $L=40$.
    \item Movie S5: Simulation of a dcABP showing a normal sliding mode inside circular confinement. Parameters used: $R=1$, $\omega=0.2$, $\alpha=-1$, $\Dr=0.01$, and $\mathrm{d}t=0.01$. 
    \item Movie S6: Simulation of a dcABP showing an anomalous sliding mode inside circular confinement. Parameters used: $R=1$, $\omega=-0.2$, $\alpha=-1$, $\Dr=0.01$, and $\mathrm{d}t=0.01$. 
    \item Movie S7: Simulation of a dcABP showing a normal sliding mode along a circular obstacle. Parameters used: $R=4$, $\omega=1$, $\alpha=-2$, $\Dr=0.01$, and $\mathrm{d}t=0.01$. 
    \item Movie S8: Simulation of two interacting dcABPs showing a spinning mode. Parameters used: $R=2.5$, $\omega=0$, $\alpha=-1$, $\Dr=0.01$, and $\mathrm{d}t=0.01$. 
    \item Movie S9: Experimental video of a doubly chiral vibrobot sliding along straight boundaries and turning both inside and outside corners. Video has been sped up 4x.
\end{itemize}

\bibliography{biblio}

@misc{kuroda2026designingtopologicaledgecurrents,
      title={Designing topological edge currents in chiral active matter}, 
      author={Yuta Kuroda and Ellen Meyberg and Gaurav Gardi and Thomas Speck and Saeed Osat},
      year={2026},
      eprint={2606.31840},
      archivePrefix={arXiv},
      primaryClass={cond-mat.soft},
      url={https://arxiv.org/abs/2606.31840}, 
}

@misc{metzger2026equation,
      title={Equation of state for the edge flow of chiral colloidal fluids}, 
      author={Jessica Metzger and Cory Hargus and Julien Tailleur and Frédéric van Wijland},
      year={2026},
      eprint={2604.18708},
      archivePrefix={arXiv},
      primaryClass={cond-mat.soft},
      url={https://arxiv.org/abs/2604.18708}, 
}

@misc{kant2025edge,
      title={Edge states, pairing, and sorting of motile chiral particles}, 
      author={Raushan Kant and Ananyo Maitra and A K Sood and Sriram Ramaswamy},
      year={2025},
      eprint={2509.00729},
      archivePrefix={arXiv},
      primaryClass={cond-mat.soft},
      url={https://arxiv.org/abs/2509.00729}, 
}

@article{baconnier2025self,
  title = {Self-aligning polar active matter},
  author = {Baconnier, Paul and Dauchot, Olivier and D\'emery, Vincent and D\"uring, Gustavo and Henkes, Silke and Huepe, Cristi\'an and Shee, Amir},
  journal = {Rev. Mod. Phys.},
  volume = {97},
  issue = {1},
  pages = {015007},
  numpages = {26},
  year = {2025},
  month = {Mar},
  publisher = {American Physical Society},
  doi = {10.1103/RevModPhys.97.015007},
  url = {https://link.aps.org/doi/10.1103/RevModPhys.97.015007}
}

@article{gnesotto2020learning,
  title={Learning the non-equilibrium dynamics of Brownian movies},
  author={Gnesotto, Federico S and Gradziuk, Grzegorz and Ronceray, Pierre and Broedersz, Chase P},
  journal={Nature Communications},
  volume={11},
  number={1},
  pages={5378},
  year={2020},
  publisher={Nature Publishing Group UK London},
  doi = {https://doi.org/10.1038/s41467-020-18796-9}
}

@article{vanZuiden2016ActiveSpinners,
  title = {Spatiotemporal order and emergent edge currents in active spinner materials},
  author = {van Zuiden, Benjamin C. and Paulose, Jayson and Irvine, William T. M. and Bartolo, Denis and Vitelli, Vincenzo},
  journal = {Proceedings of the National Academy of Sciences},
  volume = {113},
  number = {46},
  pages = {12919--12924},
  year = {2016},
  doi = {10.1073/pnas.1609572113}
}

@article{Soni2019OddFreeSurface,
  title = {The odd free surface flows of a colloidal chiral fluid},
  author = {Soni, V. and Bililign, E. S. and Magkiriadou, S. and Sacanna, S. and Bartolo, D. and Shelley, M. J. and Irvine, W. T. M.},
  journal = {Nature Physics},
  volume = {15},
  pages = {1188--1194},
  year = {2019},
  doi = {10.1038/s41567-019-0603-8}
}

@article{Yang2020RobustBoundaryFlow,
  title = {Robust boundary flow in chiral active fluid},
  author = {Yang, Xiang and Ren, Chenyang and Cheng, Kangjun and Zhang, H. P.},
  journal = {Physical Review E},
  volume = {101},
  number = {2},
  pages = {022603},
  year = {2020},
  doi = {10.1103/PhysRevE.101.022603}
}

@article{Petroff2023DensityMediatedSpin,
  title = {Density-mediated spin correlations drive edge-to-bulk flow transition in active chiral matter},
  author = {Petroff, Alexander P. and Whittington, Christopher and Kudrolli, Arshad},
  journal = {Physical Review E},
  volume = {108},
  number = {1},
  pages = {014609},
  year = {2023},
  doi = {10.1103/PhysRevE.108.014609}
}

@article{Caprini2019ActiveChiralConfinement,
  title = {Active chiral particles under confinement: surface currents and bulk accumulation phenomena},
  author = {Caprini, Lorenzo and Marini Bettolo Marconi, Umberto},
  journal = {Soft Matter},
  volume = {15},
  number = {12},
  pages = {2627--2637},
  year = {2019},
  doi = {10.1039/C8SM02492H}
}

@misc{caprini2025activethermodynamicsinertialchiral,
      title={Active thermodynamics of inertial chiral active gases: equation of state and edge currents}, 
      author={Lorenzo Caprini and Umberto Marini Bettolo Marconi and Benno Liebchen and Hartmut Löwen},
      year={2025},
      eprint={2509.05053},
      archivePrefix={arXiv},
      primaryClass={cond-mat.soft},
      url={https://arxiv.org/abs/2509.05053}, 
}

@article{Jamali2018ActiveCircularBoundaries,
  title = {Active fluids at circular boundaries: swim pressure and anomalous droplet ripening},
  author = {Jamali, Tayeb and Naji, Ali},
  journal = {Soft Matter},
  volume = {14},
  number = {23},
  pages = {4820--4834},
  year = {2018},
  doi = {10.1039/C8SM00338F}
}

@article{Halperin1982EdgeStates,
  title = {Quantized Hall conductance, current-carrying edge states, and the existence of extended states in a two-dimensional disordered potential},
  author = {Halperin, B. I.},
  journal = {Physical Review B},
  volume = {25},
  number = {4},
  pages = {2185--2190},
  year = {1982},
  doi = {10.1103/PhysRevB.25.2185}
}

@article{Buttiker1988AbsenceBackscattering,
  title = {Absence of backscattering in the quantum Hall effect in multiprobe conductors},
  author = {B{\"u}ttiker, M.},
  journal = {Physical Review B},
  volume = {38},
  number = {14},
  pages = {9375--9389},
  year = {1988},
  doi = {10.1103/PhysRevB.38.9375}
}

@article{Kummel2013CircularMotion,
  title = {Circular Motion of Asymmetric Self-Propelling Particles},
  author = {K\"ummel, Felix and ten Hagen, Borge and Wittkowski, Raphael and Buttinoni, Ivo and Eichhorn, Ralf and Volpe, Giovanni and L\"owen, Hartmut and Bechinger, Clemens},
  journal = {Phys. Rev. Lett.},
  volume = {110},
  issue = {19},
  pages = {198302},
  numpages = {5},
  year = {2013},
  month = {May},
  publisher = {American Physical Society},
  doi = {10.1103/PhysRevLett.110.198302},
  url = {https://link.aps.org/doi/10.1103/PhysRevLett.110.198302}
}

@article{Deblais2018BoundariesControl,
  title = {Boundaries control collective dynamics of inertial self-propelled robots},
  author = {Deblais, A. and Barois, T. and Guerin, T. and Delville, P. H. and Vaudaine, R. and Lintuvuori, J. S. and Boudet, J. F. and Baret, J. C. and Kellay, H.},
  journal = {Physical Review Letters},
  volume = {120},
  number = {18},
  pages = {188002},
  year = {2018},
  doi = {10.1103/PhysRevLett.120.188002}
}

@article{HasanKane2010TopologicalInsulators,
  title = {Colloquium: Topological insulators},
  author = {Hasan, M. Z. and Kane, C. L.},
  journal = {Reviews of Modern Physics},
  volume = {82},
  number = {4},
  pages = {3045--3067},
  year = {2010},
  doi = {10.1103/RevModPhys.82.3045}
}

@article{Lu2014TopologicalPhotonics,
  title = {Topological photonics},
  author = {Lu, Ling and Joannopoulos, John D. and Solja{\v{c}}i{\'c}, Marin},
  journal = {Nature Photonics},
  volume = {8},
  pages = {821--829},
  year = {2014},
  doi = {10.1038/nphoton.2014.248}
}

@article{Tang2021TopologyProtectsChiralCurrents,
  title = {Topology protects chiral edge currents in stochastic systems},
  author = {Tang, Evelyn and Agudo-Canalejo, Jaime and Golestanian, Ramin},
  journal = {Physical Review X},
  volume = {11},
  number = {3},
  pages = {031015},
  year = {2021},
  doi = {10.1103/PhysRevX.11.031015}
}

@article{AgudoCanalejo2025,
  title = {Topological phases in discrete stochastic systems},
  volume = {88},
  ISSN = {1361-6633},
  url = {http://dx.doi.org/10.1088/1361-6633/ae07fd},
  DOI = {10.1088/1361-6633/ae07fd},
  number = {10},
  journal = {Reports on Progress in Physics},
  publisher = {IOP Publishing},
  author = {Agudo-Canalejo,  Jaime and Tang,  Evelyn},
  year = {2025},
  month = Oct,
  pages = {102601}
}

@article{Shankar2022,
  title = {Topological active matter},
  volume = {4},
  ISSN = {2522-5820},
  url = {http://dx.doi.org/10.1038/s42254-022-00445-3},
  DOI = {10.1038/s42254-022-00445-3},
  number = {6},
  journal = {Nature Reviews Physics},
  publisher = {Springer Science and Business Media LLC},
  author = {Shankar,  Suraj and Souslov,  Anton and Bowick,  Mark J. and Marchetti,  M. Cristina and Vitelli,  Vincenzo},
  year = {2022},
  month = May,
  pages = {380–398}
}

@misc{wang2026edgecurrentsshapecondensates,
      title={Edge Currents Shape Condensates in Chiral Active Matter}, 
      author={Boyi Wang and Patrick Pietzonka and Frank Jülicher},
      year={2026},
      eprint={2603.20064},
      archivePrefix={arXiv},
      primaryClass={cond-mat.stat-mech},
      url={https://arxiv.org/abs/2603.20064}, 
}

@article{vanTeeffelen2008,
  title = {Dynamics of a Brownian circle swimmer},
  author = {van Teeffelen, Sven and L\"owen, Hartmut},
  journal = {Phys. Rev. E},
  volume = {78},
  issue = {2},
  pages = {020101(R)},
  numpages = {4},
  year = {2008},
  month = {Aug},
  publisher = {American Physical Society},
  doi = {10.1103/PhysRevE.78.020101},
  url = {https://link.aps.org/doi/10.1103/PhysRevE.78.020101}
}

@article{teeffelen2009clockwise,
    author = {van Teeffelen, Sven and Zimmermann, Urs and Löwen, Hartmut},
    title = {Clockwise-directional circle swimmer moves counter-clockwise in Petri dish- and ring-like confinements},
    journal = {Soft Matter},
    volume = {5},
    number = {22},
    pages = {4510-4519},
    year = {2009},
    month = {11},
    issn = {1744-683X},
    doi = {10.1039/b911365g},
    url = {https://doi.org/10.1039/b911365g}
}

@article{carrillomora2025depinningactivatedmotionchiral,
  title = {Depinning and activated motion of chiral self-propelled particles},
  author = {Carrillo-Mora, Juan Pablo and Garc\'es, Adri\`a and Levis, Demian},
  journal = {Phys. Rev. E},
  volume = {112},
  issue = {6},
  pages = {065417},
  numpages = {6},
  year = {2025},
  month = {Dec},
  publisher = {American Physical Society},
  doi = {10.1103/7zmp-z893},
  url = {https://link.aps.org/doi/10.1103/7zmp-z893}
}

@article{PhysRevX.12.041017,
  title = {Chiral Edge Current in Nematic Cell Monolayers},
  author = {Yashunsky, V. and Pearce, D. J. G. and Blanch-Mercader, C. and Ascione, F. and Silberzan, P. and Giomi, L.},
  journal = {Phys. Rev. X},
  volume = {12},
  issue = {4},
  pages = {041017},
  numpages = {11},
  year = {2022},
  month = {Nov},
  publisher = {American Physical Society},
  doi = {10.1103/PhysRevX.12.041017},
  url = {https://link.aps.org/doi/10.1103/PhysRevX.12.041017}
}

@article{PhysRevX.14.041006,
  title = {Robust Edge Flows in Swarming Bacterial Colonies},
  author = {Li, He and Chat\'e, Hugues and Sano, Masaki and Shi, Xia-qing and Zhang, H. P.},
  journal = {Phys. Rev. X},
  volume = {14},
  issue = {4},
  pages = {041006},
  numpages = {15},
  year = {2024},
  month = {Oct},
  publisher = {American Physical Society},
  doi = {10.1103/PhysRevX.14.041006},
  url = {https://link.aps.org/doi/10.1103/PhysRevX.14.041006}
}

@article{Cooper1997Thermoelectric,
  title = {Thermoelectric response of an interacting two-dimensional electron gas in a quantizing magnetic field},
  author = {Cooper, N. R. and Halperin, B. I. and Ruzin, I. M.},
  journal = {Physical Review B},
  volume = {55},
  number = {4},
  pages = {2344--2359},
  year = {1997},
  doi = {10.1103/PhysRevB.55.2344}
}

@article{PhysRevLett.107.236601,
  title = {Energy Magnetization and the Thermal Hall Effect},
  author = {Qin, Tao and Niu, Qian and Shi, Junren},
  journal = {Phys. Rev. Lett.},
  volume = {107},
  issue = {23},
  pages = {236601},
  numpages = {5},
  year = {2011},
  month = {Nov},
  publisher = {American Physical Society},
  doi = {10.1103/PhysRevLett.107.236601},
  url = {https://link.aps.org/doi/10.1103/PhysRevLett.107.236601}
}

@article{274x-4vgn,
  title = {Free chiral self-propelled robots compared to active Brownian circle swimmers},
  author = {Kiechl, Thomas and Altshuler, Amy and L\"uders, Anton and Roichman, Yael and Franosch, Thomas},
  journal = {Phys. Rev. E},
  volume = {113},
  issue = {4},
  pages = {045409},
  numpages = {13},
  year = {2026},
  month = {Apr},
  publisher = {American Physical Society},
  doi = {10.1103/274x-4vgn},
  url = {https://link.aps.org/doi/10.1103/274x-4vgn}
}

@article{lauga2006swimming,
title = {Swimming in Circles: Motion of Bacteria near Solid Boundaries},
journal = {Biophysical Journal},
volume = {90},
number = {2},
pages = {400-412},
year = {2006},
issn = {0006-3495},
doi = {https://doi.org/10.1529/biophysj.105.069401},
url = {https://www.sciencedirect.com/science/article/pii/S0006349506722214},
author = {Eric Lauga and Willow R. DiLuzio and George M. Whitesides and Howard A. Stone}
}

@article{lettermann2026chirality,
  title={Chirality of malaria parasites determines their motion patterns},
  author={Lettermann, Leon and Singer, Mirko and Steinbr{\"u}ck, Smilla and Ziebert, Falko and Kanatani, Sachie and Sinnis, Photini and Frischknecht, Friedrich and Schwarz, Ulrich S},
  journal={Nature Physics},
  volume={22},
  number={1},
  pages={112--122},
  year={2026},
  publisher={Nature Publishing Group UK London},
  doi = {https://doi.org/10.1038/s41567-025-03096-0}
}

@article{PhysRevE.82.015304,
  title = {Self-assembled autonomous runners and tumblers},
  author = {Ebbens, Stephen and Jones, Richard A. L. and Ryan, Anthony J. and Golestanian, Ramin and Howse, Jonathan R.},
  journal = {Phys. Rev. E},
  volume = {82},
  issue = {1},
  pages = {015304(R)},
  numpages = {4},
  year = {2010},
  month = {Jul},
  publisher = {American Physical Society},
  doi = {10.1103/PhysRevE.82.015304},
  url = {https://link.aps.org/doi/10.1103/PhysRevE.82.015304}
}

@article{brambilla2013swarm,
  title={Swarm robotics: a review from the swarm engineering perspective},
  author={Brambilla, Manuele and Ferrante, Eliseo and Birattari, Mauro and Dorigo, Marco},
  journal={Swarm Intelligence},
  volume={7},
  number={1},
  pages={1--41},
  year={2013},
  publisher={Springer},
  doi = {https://doi.org/10.1007/s11721-012-0075-2}
}

\end{document}